\documentclass[journal]{IEEEtran}

\usepackage[T1]{fontenc}
\usepackage[latin9]{inputenc}
\usepackage{color}
\usepackage{float}
\usepackage{bbding}

\usepackage{cite}
\usepackage{array} 
\usepackage{multirow}
\usepackage{amsmath,amssymb,amsfonts}
\usepackage{algorithmic}
\usepackage{graphicx}
\usepackage{textcomp}
\usepackage{xcolor}
\usepackage{pgfplots}
\usepackage[keeplastbox]{flushend}

\providecommand{\tabularnewline}{\\}
\floatstyle{ruled}
\newfloat{algorithm}{tbp}{loa}
\providecommand{\algorithmname}{Algorithm}
\floatname{algorithm}{\protect\algorithmname}

\def\BibTeX{{\rm B\kern-.05em{\sc i\kern-.025em b}\kern-.08em
    T\kern-.1667em\lower.7ex\hbox{E}\kern-.125emX}}
\usepackage{soul}

\begin{document}

\title{AMOGA: A Static-Dynamic Model Generation Strategy for Mobile Apps Testing}

\author{Ibrahim-Anka Salihu,
        Rosziati Ibrahim,
        Bestoun S. Ahmed*,
        Kamal Z. Zamli, and 
        Asmau Usman

\thanks{I. A. Salihu: Department of Software Engineering Faculty of Computer Science and Information Technology Universiti Tun Hussein Onn Malaysia 86400 Batu Pahat, Johor, Malaysia, email: isankah3@gmail.com}  

\thanks{R. Ibrahim: Department of Software Engineering Faculty of Computer Science and Information Technology Universiti Tun Hussein Onn Malaysia 86400 Batu Pahat, Johor, Malaysia, email: rosziati@uthm.edu.my}
        
\thanks{B. Ahmed is with the Department of Mathematics and Computer Science, Karlstad University, Karlstad, Sweden
email: bestoun@kau.se}

\thanks{K. Z. Zamli: Faculty of Computer Systems and Software Engineering, Universiti Malaysia Pahang 26300 Gambang, Pahang, Malaysia, email: kamalz@ump.edu.my}

\thanks{A. Usman: Department of Software Engineering Faculty of Computer Science and Information Technology Universiti Tun Hussein Onn Malaysia 86400 Batu Pahat, Johor, Malaysia, email: gi160007@siswa.uthm.edu.my}}

\maketitle

\begin{abstract}
In the past few years, mobile devices have been increasingly replacing traditional computers as their capabilities such as CPU computation, memory, RAM size, and many more, are being enhanced almost to the level of conventional computers. These capabilities are being exploited by mobile apps developers to produce apps that offer more functionalities and optimized performance. To ensure acceptable quality and to meet their specifications (e.g., design), mobile apps need to be tested thoroughly. As the testing process is often tedious, test automation can be the key to alleviating such laborious activities. In the context of the Android-based mobile apps, researchers and practitioners have proposed many approaches to automate the testing process mainly on the creation of the test suite. Although useful, most existing approaches rely on reverse engineering a model of the application under test for test case creation. Often, such approaches exhibit a lack of comprehensiveness as the application model does not capture the dynamic behavior of the applications extensively due to the incompleteness of reverse engineering approaches. To address this issue, this paper proposes AMOGA, a strategy that uses a hybrid, static-dynamic approach for generating user interface model from mobile apps for model-based testing. AMOGA implements a novel crawling technique that uses the event list of UI element associated with each event to dynamically exercise the events ordering at the run-time to explore the applications' behavior. An experimental evaluation was performed to assess the effectiveness of our strategy by measuring the code coverage and the fault detection capability through the use of mutation testing concept. Results of the experimental assessment showed that AMOGA represents an alternative approach for model-based testing of mobile apps by generating comprehensive models to improve the coverage of the applications. The strategy proved its effectiveness by achieving high code coverage and mutation score for different applications.
\end{abstract}


\IEEEpeerreviewmaketitle

\section{Introduction\label{sec:Introduction}}

Smartphones and tablets have dominated the global computing trend in recent years \cite{Gartner}. A recent study indicated that a total of 383,504 smartphones were sold to users worldwide in the first quarter of 2018 and Android OS is leading with 85.9\% market share \cite{Gartner}; whereas, the personal computers (laptop and desktop) sold to users worldwide in the same period are 61,686 \cite{GartnerWebSite}. This indicates that smartphones are gradually replacing laptops and desktop computers for performing many computational tasks such as access to email, internet surfing, spreadsheet generation/editing, word processing, and presentation making/editing. This shift has significantly changed the computing landscape. The popularity of these devices has brought an increase in the development of mobile applications (apps) to deal with the computational needs of their users \cite{nayebi2012state}. Mobile apps development has a significant impact from both economic and social perspectives. It has generated revenue of \$86 billion in 2017, and a recent report estimated that the global applications business would be worth \$110 billion in 2018 and \$189 by 2020 \cite{TechCrunchWebSite,StatistaWebSite}.

The increase in complexity of mobile apps due to the increase in capacity, structure, and functionalities has brought several challenges for the software engineering researchers such as determining/deriving application's behaviors and testing them \cite{yang2013grey,Yan2015static}. Consequently, there is a demand for software engineering techniques and tools to support the analysis and testing task for mobile apps \cite{dehlinger2011mobile,janicki2012obstacles,muccini2012software}. Testing can play a significant role in assessing and improving the quality of software systems \cite{osterweil1996strategic,amalfitano2011gui}. With the recent improvements and complexity of mobile apps, manual testing is no longer sufficient because it is often tedious, error-prone, and achieves poor coverage of an application's behavior \cite{yang2013grey,salva2016model}. For example, when there is a need to cover a large number of combinations of usage scenarios, it can be tedious to manually enter data, swipe the screen, and click on buttons.

Model-based testing (MBT) is a popular approach for test automation where a model of the application under test (AUT) is build to derive the test input automatically \cite{banerjee2013graphical}. It provides a notable improvement to conventional scripted testing by enhancing the creation of test scripts and test coverage of an application \cite{lu2012automated,kull2012automatic,nguyen2014guitar}. However, due to the use of agile development processes by most developers in which the requirements and implementation are iterating rapidly from a version to another, modeling in such a situation is difficult \cite{kull2012automatic,wasserman2010software}. Hence, mobile apps model is often not available or of insufficient quality \cite{baek2016automated}. The model can be generated from the application's documentation \cite{utting2012taxonomy} or sequences of actions observed in an application \cite{kull2012automatic}. The reverse engineering approach has been used recently to generate the model automatically \cite{kull2012automatic,aho2015making}. To benefit from the MBT approach, there is a demand for techniques/tools to aid automated model generation from the mobile apps. However, building these models fully automatically for the Android apps have several challenges such as exploring system events (e.g., events due to the incoming call) and some events that are only visible by toggling the visibility-property of a panel \cite{amalfitano2012using,grilo2010reverse}.

In the context of UI testing, many reverse engineering tools for the automated model generation from mobile apps have emerged. Most of these tools are typically based on dynamic approaches where an application is dynamically analyzed at the run-time to extract information. For example, Android GUITAR \cite{SourceforgeGuitarPage}, Android GUI Ripper \cite{amalfitano2012using}, MCrawlT \cite{salva2016model}, and test automation system \cite{tao2016building} are all based on this approach. However, the information extracted by a pure dynamic approach is incomplete due to the inability to explore infeasible paths (e.g., windows that require a password) and providing user inputs \cite{morgado2012dynamic,kull2012automatic}. As such the models generated by these tools are incomplete due to the limitations of pure dynamic analysis \cite{silva2013combining,salihu2016comparative}. Tools that combine both static (analysis of bytecode/source code to extract valuable information) and dynamic (analyzing application at the run time) approaches were proposed recently to improve the coverage and the quality of the generated models from the mobile apps such as Orbit \cite{yang2013grey} and A3E \cite{azim2013targeted}. Nonetheless, the models generated by these tools are incomplete. One of the limitations of these approaches is that the static analysis employed by them is less comprehensive as it does not capture menus and dialog, it does not consider the UI effects of event handlers and the triggered callbacks \cite{Shengqian2015static,yang2015static}.

Addressing the issues raised above, this paper proposes a hybrid approach, by combining static and dynamic technique. As the name implies, the static technique extracts the mobile app\textquoteright s events statically by analyzing the corresponding bytecodes. Meanwhile, the dynamic technique matches the event list of UI element associated with each event to dynamically exercise the events ordering at the run-time to explore the applications\textquoteright{} behavior. The contributions of this work can be summarized as follows: 
\begin{itemize}
\item We propose a novel dynamic crawling algorithm for exploring a mobile application and constructing the interaction model of the UI. The new crawler reduces the crawling and model construction time by using an enhanced search algorithm.  

\item Unlike the other approach, we also propose a groundbreaking approach towards completing and supporting the dynamically constructed model (i.e., using the crawler) by refining it using a static analysis algorithm for producing the mobile apps\textquoteright{} events.

\item We developed an Automated Model Generator for Android apps (AMOGA), a tool for automated UI model generation from mobile applications. \item We evaluated the effectiveness of the proposed novel approach on different real worlds case studies of Android apps and compared the results with the current state-of-the-art tools and algorithms. We showed the effectiveness of AMOGA in terms of code coverage and fault detection ability. 
\end{itemize}

The rest of this paper is organized as follows. Section \ref{sec:Background} discusses the background. Section \ref{sec:Motivating-Example} presents a motivating example. Section \ref{sec:The-Proposed-Approach} presents the proposed hybrid approach. Section \ref{sec:Experimetal-Evaluation} discussed the results of the experimental evaluation. Section \ref{sec:Related-Work} discusses the related works. Finally, section \ref{sec:Conclusion} concludes the paper.

\section{Background\label{sec:Background}}

The popular tool categories in the literature for automated test cases generation and execution in the context of mobile apps are script-based, capture/replay, random walk, systematic exploration, and model-based \cite{choudhary2015automated,banerjee2013graphical}. The script-based technique requires writing test cases manually to automatically interact with the GUI using scripting languages that programmatically control the GUI. The languages utilize the JUnit framework \cite{JUnitWebSite}, a tool developed for unit testing. Capture/replay tools are interactive tools that reduce the burden of manual scripting through the provision of interactive tool support. They provide a mechanism that allows the tester to record interactions with the UI and save the interactions as test cases that can be replayed automatically. An example of capture/replay tool is Selenium IDE \cite{SeleniumWebSite}. Random walk tools explore the UI and execute all events encountered in sequence. As the tools do not generate test cases, they cannot replay the exact sequences encountered. An example of such tool is Android Monkey \cite{GoogleCodeWebSite}. Systematic exploration tools use more advanced techniques, such as symbolic execution, to guide the exploration upon providing specific inputs. Symbolic execution analyses the code of a program and automatically generates test data for it. An example of this tool is A3E \cite{azim2013targeted}.

MBT is a popular area of research in recent years. It can enhance the creation of test scripts and test coverage of an application \cite{nguyen2014guitar}. As the model depicts an abstract representation of the expected behavior, the test scripts that are generated from such model can check conformance of the implementation with the expected behavior \cite{kull2012automatic}. Using a model to depict the behavior of a software system has been proven to be of significant advantage. Nonetheless, building the model is one of the crucial steps in MBT when the model is not available. It can be constructed manually as in \cite{takala2011experiences}, or using automatic modeling techniques \cite{memon2003gui,paiva2007reverse}. However, constructing the model manually is tedious, error-prone and time-consuming \cite{yang2013grey} as it requires careful inspection of the application to represent the GUI at design and implementation levels. Automated model generation is a reverse engineering task that involves extracting the design artifacts and deriving abstractions of an application. Hence, the quality of a model with respect to its level of abstraction depends on the amount of information captured. 

Several model reverse engineering techniques/tools were proposed for automated testing of Android apps over the last decade. Most of these tools are pure black-box techniques that perform dynamic analysis of applications. A few of those tools are based on the gray-box (hybrid) technique.

The black-box technique is a method of testing where the tester examines the behavior of an application without knowing the app's code/internal structure \cite{limaye2009software}. The focus is mainly on the outputs generated in response to the selected inputs and execution conditions. In this technique, test input can be obtained by dynamically analyzing an application at run-time \cite{limaye2009software} or from the external descriptions of the application, including specifications, requirements and design parameters \cite{lin2014accuracy,liu2009covering}. It is particularly suited for extracting information about the UI\textquoteright s external behavior \cite{silva2013combining}. One of the most challenging issues in dynamic reverse engineering is how events are found and fired in controlling the model exploration. Also, the inability to explore certain UI due to the presence of infeasible paths such as those that require user inputs and modal UI (dialog box) that can sometimes be visible or invisible \cite{silva2013combining,kull2012automatic}. Hence, the extracted information about the behavior of the application could be inaccurate and incomplete which affects the quality of the generated model.

White-box technique (otherwise known as code-based testing) is based on the static analysis of the code to test the internal structures of an application. White-box technique involves the analysis of the application\textquoteright s code without executing the application itself to generate a human-readable form of representation \cite{limaye2009software}. It is well suited for extracting information about the internal structure of the system and dependencies among the structural elements such as classes, methods, and variables information \cite{silva2013combining}. Furthermore, it can retrieve more accurate information from an application. Nonetheless, the dynamic object-oriented nature of the UI applications makes it difficult or even impossible to generate comprehensive information about the behavior of the application (e.g., behaviors of the dynamically created UI widgets) by just analyzing their source code \cite{kull2012automatic}. 

Gray-box technique is a combination of the black-box and white-box technique \cite{choudhary2015automated}. In this technique, the tester performs a static analysis (white-box) of the application's code to acquire the inputs that can be used for the dynamic analysis (black-box). Nowadays, this approach otherwise referred to as the hybrid approach. Recently, the hybrid approach has been the focus of researchers in the area of UI reverse engineering, particularly for the Android apps \cite{aho2014murphy}. The hybrid approach can provide enhancement in terms of the scope, completeness, and precision of reverse engineering as it exploits the capabilities of both static and dynamic approaches while trying to maximize the quality of the extracted information \cite{morgado2012dynamic}.

\section{Motivating Example\label{sec:Motivating-Example}}

\begin{figure*}
\begin{centering}
\includegraphics[scale=0.4]{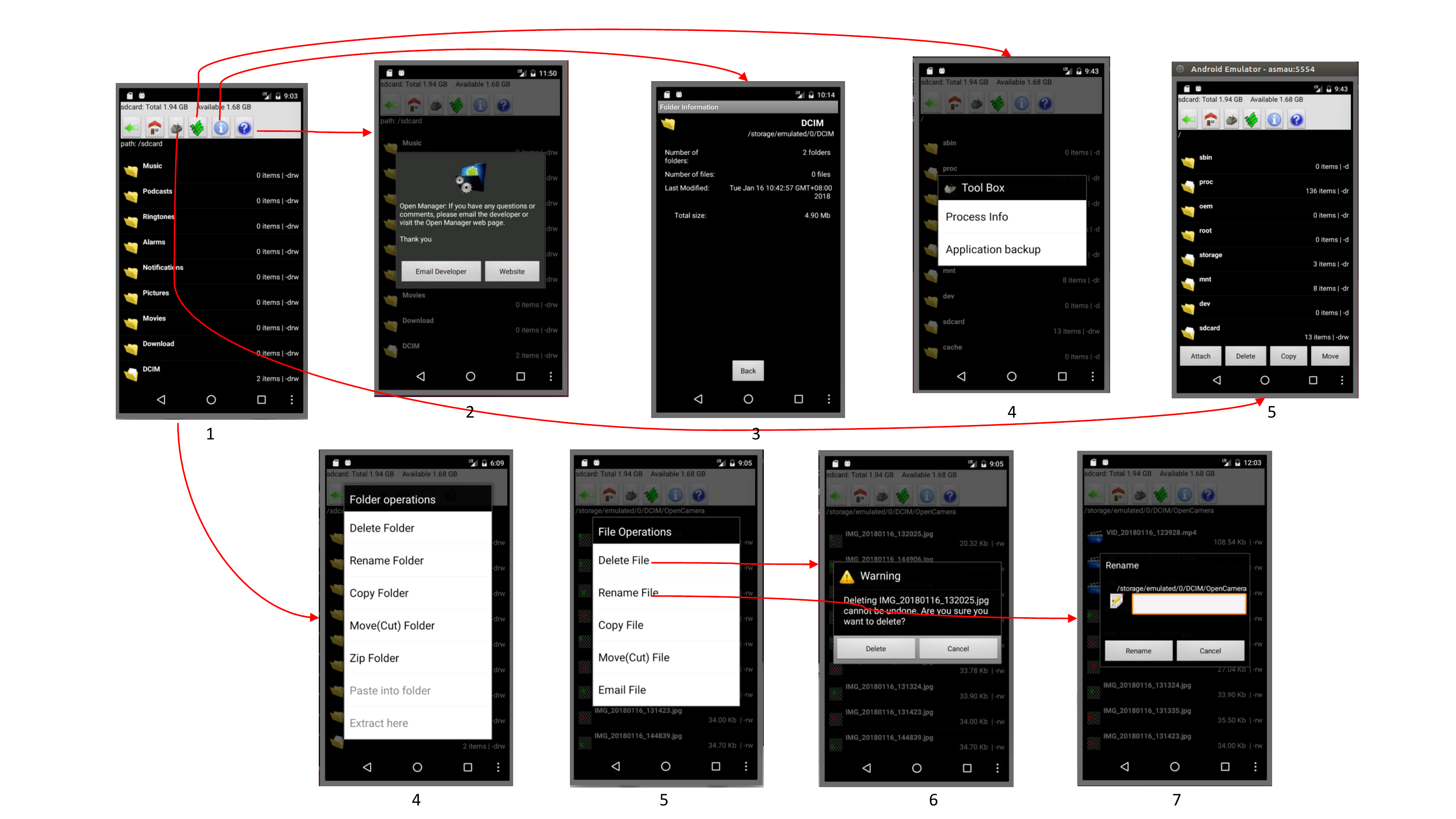}
\par\end{centering}
\caption{Example derived from OpenManager app\label{fig:Example-derived-from}}
\end{figure*}

Figure \ref{fig:Example-derived-from} shows a simple example derived from an Android app called OpenManager. OpenManager is an open-source file manager app that allows you to browse your device, create directories, rename, copy, move, and delete files. It consists of thirteen windows (five Activities and eight dialogs). The main window displays directories currently in the device. Clicking on a directory displays its contents. The help button displays a dialog (image 2) with two options, Email Developer for the user to ask the developer any question about the application and Website option that directs to the website of the application. The directoryInfo button displays a new window with the current directory information (image 3). A click on the manage button displays a new window (image 4). The window is an instance of the AlertDialog that is used to show two selectable items (e.g., the running process info, or to backup applications to the SD card). The multi-select button displays a window that is associated with the main window with changes in its visual representation (image 5). It consists of the attach button for attaching a file, delete button to delete a file, copy and move button for copying or to delete a file. Finally, the menu is opened by clicking the menu button from any window (not shown in the figure). The new directory option in the menu leads to a dialog for typing the new directory name. The search option takes the users to another dialog for typing name of the file to search. The settings option on the menu directs the user to a window with five settings options. The quit option in the menu closes the application.

The other windows 4-7 are dialogs that are not directly accessible via a simple button click. A long click on any directory opens a dialog, \textit{folder operations} (image 4), with several selectable options. A long click on any file in a directory leads to a dialog, file operations (image 5), with several selectable options. Clicking on the delete file option in the file operation dialog takes the user to a warning dialog (image 6) to confirm the request and clicking the rename option leads the user to a dialog to type the new name. The long click events are examples of events that are typically inaccessible by the dynamic approach. The dialogs and menus are other essential applications' windows that affect the visible state and possible run-time behavior of applications. These events are not handled by the other hybrid approaches/tools as discussed in section \ref{sec:Related-Work}. 

\section{The Proposed Approach\label{sec:The-Proposed-Approach}}

\begin{figure*}
\begin{centering}
\includegraphics[scale=0.4]{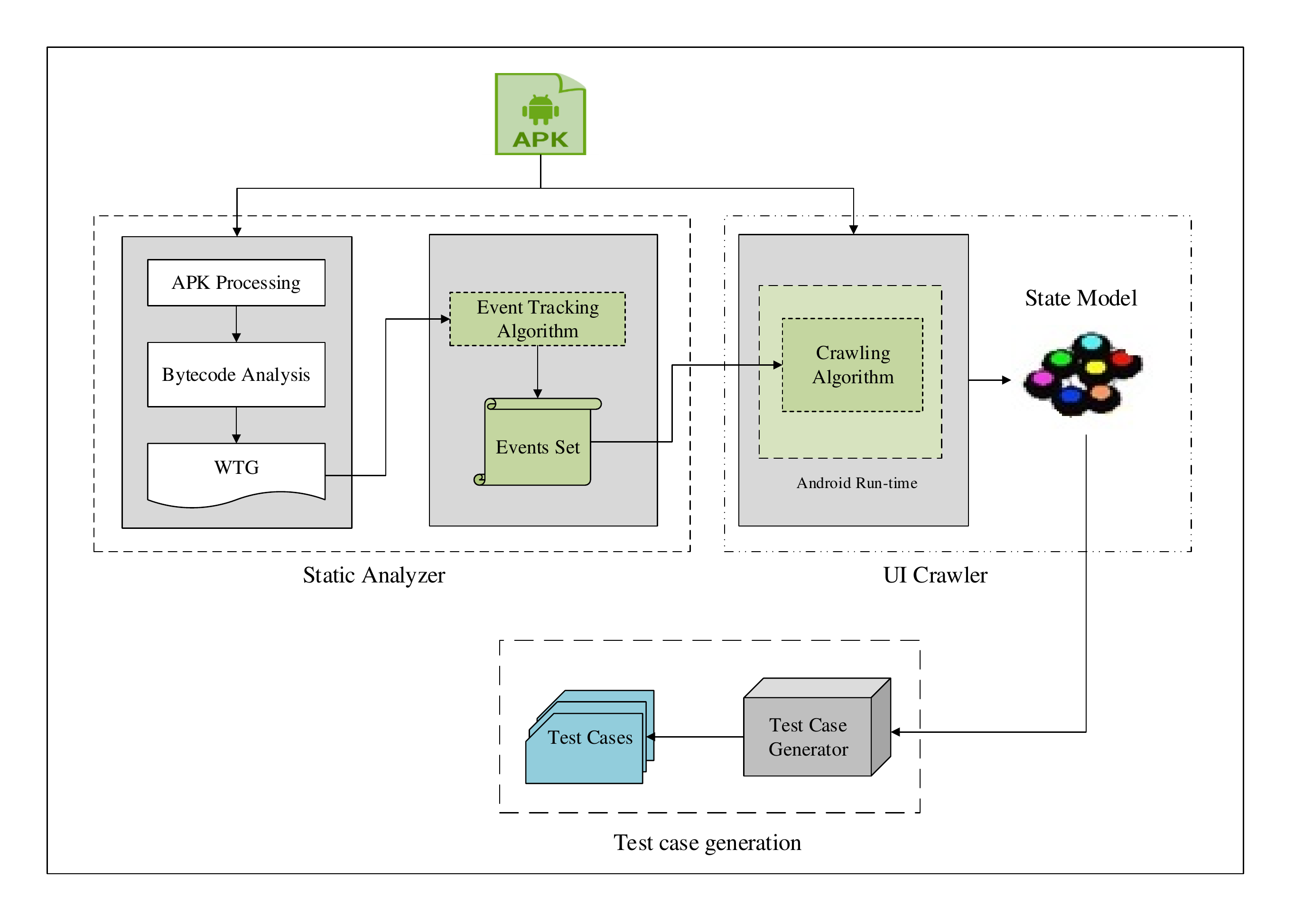}
\par\end{centering}
\caption{Framework of AMOGA \label{fig:Framework-of-AMOGA}}
\end{figure*}

We now describe the hybrid approach for reverse engineering model of a given mobile app. As discussed in the previous section, one of the significant challenges in testing mobile apps is how to generate input events that are used to control the exploration. Our approach performs a static analysis of the mobile app\textquoteright s bytecode to extract a set of events supported by the UIs which can be used as input for the dynamic analysis. This step is followed by a dynamic crawler, whose primary goal is to systematically fire the extracted events on the running application to explore and reverse engineer a model of the application. The framework of the proposed approach is shown in Figure \ref{fig:Framework-of-AMOGA}. The following subsections elaborate these steps in detail. 

\subsection{Events Tracking Algorithm\label{subsec:Events-Tracking-Algorithm}}

Exploring a mobile app for model reverse engineering requires knowledge of the precise set of events supported by the application. A common practice is the use of dynamic analysis to rip this information and use it for the exploration. Most state-of-the-art tools such as MobiGUITAR \cite{Amalfitano2015} are based on the dynamic approach that analyzes the application\textquoteright s UI at run-time to extract and create a task list that can be used as input in controlling the model exploration. However, with the limitations of dynamic analysis as discussed in Section \ref{sec:Background}, several researchers believed that using static analysis to generate meaningful input for the dynamic exploration can ensure the generation of a high-quality model \cite{yang2013testing}.

The control flow in Android is non-standard as in traditional Java applications but centered around callbacks invoke by the Android framework as the users trigger actions \cite{yang2013grey,azim2013targeted}. The callback methods can significantly affect the life-cycle and consequently state of an application. Thus, analyzing callbacks is essential to ensure comprehensive coverage of application\textquoteright s state. AMOGA performs a callback control-flow analysis on mobile apps. 

The common types of system events triggered by the Android system are the calls to launch-activity (e.g., the startActivity calls) which uses an intent object to specify the target activity, interrupt due to an incoming phone call and screen rotation. Several techniques for analysis of Android intent objects are available \cite{fuchs2009scandroid,feng2014apposcopy,huang2014asdroid}. Our static analysis comprises of an intent analysis which is derived from \cite{octeau2013effective}. The intent analysis is combined with the tracking of calls to launch window and the modeling of window termination calls. The system events are generated through the intent analysis. The analysis is represented in the form a Windows Transition Graph (WTG).

The WTG is a graph with nodes corresponding to windows and edges representing windows transitions. Each edge is assigned a label indicating the sequence of execution. The events tracking algorithm in Algorithm \ref{alg:Events-Tracking-1} is applied on the WTG to traverse the graph and create a set of events that can be used to explore an application dynamically. Here, the events sequence is crucial because the availability of some states defends the existence of other states. Unlike other approaches, the events tracking algorithm in AMOGA generates the events set based on their sequence of execution. This feature represents an excellent addition to the current state of the art with these tools. For example, the sequence of events is not considered in the exploration of the ORBIT tool.

\begin{algorithm*}
\caption{\label{alg:Events-Tracking-1}Events Tracking}

\textbf{Input:} WTG $g=(w,p)$

\textbf{Output:} \textit{eS} = eventSet

1. Procedure EventsTracking(g)

2. Initialization: Priority queue(PQ), visited set(vSet)

3. Enqueue intial window(W, 0)

4. \textbf{while} $PQ\neq empty$ 

5. $\;$$\;$$dequeueCurrFrom\:PQ$

6. $\;$$\;$$addCurrToVisitedSet$ 

7. $\;$$\;$\textbf{ foreach }curr neighbors, n \textbf{do} //paths
connecting to other nodes

8. $\;$$\;$$\;$ $enqueue\;n$ onto PQ //enqueue n with its priority

9. $\;$$\;$ $\;$ eventSet $eS\leftarrow getAll\;n\;from\;vSet$
//starting with n with min priority

10. $\;$$\;$$\,$\textbf{ foreach} event $e\in eS$ \textbf{do}

11. $\;$$\;$$\;$$\;$ $sWindow\leftarrow getSourceWindow(g)$

12. $\;$$\;$$\;$$\;$ $mTrigger\leftarrow getTriggerMethod(e)$

13. $\;$$\;$$\;$ $\;$ \textbf{foreach$mTrigger\in mTriggerSet$}
\textbf{do}

14. $\;$$\;$$\;$ $\;$$\;$ $v\leftarrow getWidget(t)$

15. $\;$$\;$$\;$ $\;$$\;$ $id\leftarrow getParameters(v)$

16. $\;$$\;$$\;$ $\;$$\;$ $e.S.add(e,id)$

17. $\;$$\;$$\;$ $\;$ \textbf{end}

18. $\;$$\;$$\;$ \textbf{end}

19. $\;$$\;$\textbf{end} 

20. \textbf{end} 
\end{algorithm*}

Algorithm \ref{alg:Events-Tracking-1} describes the steps of the events tracking. The algorithm starts with the WTG as input. First, we initialize a priority queue \textit{(PQ)} and begin by adding the initial node (a node with no incoming edges) to the \textit{PQ}. The priority is computed using the level (weight) on the edges of the graph. The edges with lower sequence number are given high priority and are executed before those with a higher sequence number. The algorithm continues the traversal by removing an element with the minimum value (lower sequence number) from the\textit{ PQ} (Line 5). By using a priority queue to guide the traversal, it is possible to control the traversal of the paths base on their sequence in the \textit{PQ}. The algorithm proceeds to find all neighbors, \textit{n} of the current node and add them to the \textit{PQ} with their sequence \textit{S} until all neighbors are explored (Lines 7-8). Then, the algorithm generates the eventSet from the visited set based on their sequence (Line 9). For each extracted event,\textit{ e}, the algorithm locates the source UI (node) of the event and the trigger methods (Lines 10-12). Finally, the algorithm slice back to get the widgets that are registered to the trigger methods with the view ids and extract them (Lines 13-16).

\subsection{Crawling Algorithm \label{subsec:Crawling-Algorithm}}

Automated model exploration tools use a ripper or crawler that typically require information of the set of application\textquoteright s UI widgets supporting events, such as clicks, and what events are supported explicitly by each widget \cite{choudhary2015automated}. This information is used to create a task list that can be used in triggering events to explore application's interfaces at run-time. The exploration is usually implemented with crawlers that are guided by the Depth First Search (DFS) algorithm. A DFS needs to back-track several times to make sure all paths (edges) in a graph are covered. However, when the graph to be explored is large, redundancy becomes more prevalent causing an increase in computational time. Breadth First Search (BFS) algorithm can provide an improvement over DFS. Nonetheless, both the classical DFS and BFS are based on stack data structure and they are not suitable for a weighted graph as in our case. ORBIT tool minimized this issue with their ``forwardCrawlfromState'', a modified DFS algorithm which visits the nodes and identifies a state with open action and continues crawling until it reaches a state with no open action. It will then backtrack until it reaches another open state. However, DFS does not account for weighted (the event sequence) edges as we have in our case. This is essential in selecting which event to fire next during the exploration. To efficiently crawl an application UI, our crawler implements Dijkstra's algorithm which is an enhanced form of BFS for search optimization \cite{kleinberg2006algorithm}. The main aim of the algorithm is to find the shortest path to explore the UI. It generates the shortest sequence of events based on the received \textit{eS} from the static analyzer to crawl a given application. Algorithm \ref{alg:App-crawling-1} provides a detailed description of the application crawling.

\begin{algorithm*}
\caption{\label{alg:App-crawling-1}App crawling}

\textbf{Input:} A: app under test, eS: event set

\textbf{Output: }M: generated model

1. Initialize $M\leftarrow\emptyset$; $eS\leftarrow getEventsSet$,
$vS:visited\;HashSet$

2. currentState\textbf{ }$s_{c}\leftarrow pushInitialUI$

3. \textbf{while }$eS\neq empty$ 

4. $\;$$\;$$e\leftarrow getEventFrom\;eS$

5. $\;$$\;$ pick the UI widget to fire

6. $\;$$\;$ $fireAction$ on $e$

7. $\;$$\;$ $analyseNextState$

8. $\;$$\;$ \textbf{if} $=newState$ \textbf{then}

9. $\;$$\;$$\;$$\;$ $addToModel$

10. $\;$$\;$$\;$ $updateVisitedStates,\:vS$

11. $\;$$\;$\textbf{endif}

12. $\;$$\;$$e_{n}\leftarrow getNextEventToExplore$

13. $\;$$\;$\textbf{if} $e_{n}\neq currUI$

14. $\;$ $\;$$\;$ $\mathbf{B{\scriptstyle ACKTRACK}P{\scriptstyle ROCEDURE}}$

15. $\;$ $\;$$\;$ $backtrackToPreviousUI$

16. $\;$ $\;$$\;$ $analyzeCurrUI$

17. $\;$ $\;$$\;$ \textbf{if} $e_{n}$ is reachable from curr UI
\textbf{then}

18. $\;$ $\;$$\;$$\;$$\;$ $fireAction$ on $e$

19. $\;$ $\;$$\;$$\;$$\;$ $updateModel$ //add newState to model

20. $\;$ $\;$$\;$\textbf{ else} 

21. $\;$ $\;$$\;$$\;$$\;$ goto next UI

22. $\;$ $\;$$\;$$\;$$\;$ $updateModel$

23. $\;$ $\;$$\;$ \textbf{end} 

24. $\;$$\;$ \textbf{end}

25. \textbf{end}
\end{algorithm*}

The crawling algorithm performs four main decisions repeatedly until all events in the \textit{eS} are explored. The first step is selecting the next event to explore and fire action on the widget (Line 5). Then it checks if the next \textit{UI} is a new state (Line 8). If it is a new state, it will be added to the model and the crawling continue. Then, the algorithm checks whether the next event to explore can be reached from the current \textit{UI} (Line 13). If the next event is unreachable from the current \textit{UI}, the backtrack mechanism is called to backtrack to the previous \textit{UI} until the next event is reached. Finally, the algorithm will fire the action if the next event \textit{e} is reachable from the current \textit{UI} (Line 18). The model is updated and the function to get the next event from the \textit{eS} is used to continue crawling.

\begin{figure*}
\begin{centering}
\includegraphics[scale=0.4]{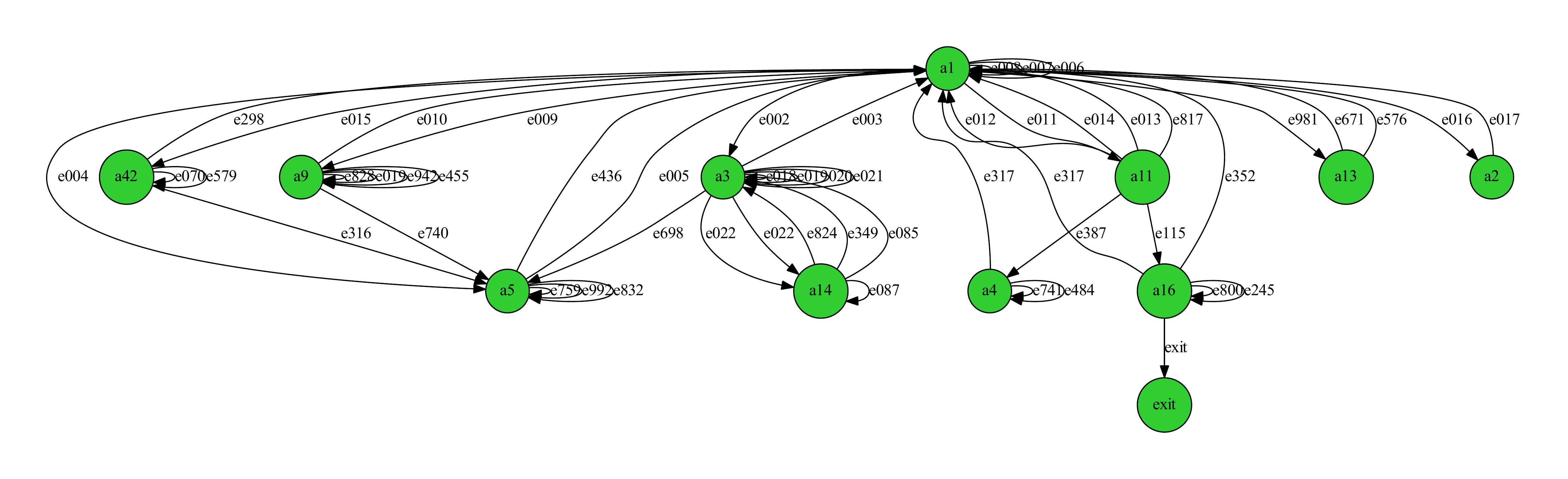}
\par\end{centering}
\caption{\label{fig:State-model-for-Tomdroid}Example of model generated for
Tomdroid app}
\end{figure*}

We modeled the UI behavior of a mobile app as an FSM. The model consists of nodes representing \textit{UI }states and edges representing events and interactions. Each input event may trigger an abstract state transition in the machine. The state machine can be used to generate test cases for testing an application. The FSM maps events and related conditions to a list of \textit{UI} actions references. Figure \ref{fig:State-model-for-Tomdroid} shows an example of the state model generated for the Tomdroid app.

AMOGA reduces the model crawling time by utilizing an enhanced search algorithm, the Dijkstra\textquoteright s algorithm. The algorithm enables AMOGA to quickly crawl the shortest \textit{UI}-path to explore apps\textit{ }states and continue the crawling process from the lastly discovered app state to discover and explore new undiscovered app\textquoteright s state. In comparison to other tools (SwiftHand, MCrawlT, MobiGUITAR) which have a time complexity of $O(s^{n})$, where$s$ is the number of subsequent paths originating from a particular app \textit{UI}, AMOGA achieved the lowest crawling time. Its time complexity is $O(n)$ in the worst-case, and it is linear to the number of an app \textit{UIs} $(n)$ on the path leading to a target app \textit{UI} (the depth).

\subsection{Test case Generation\label{subsec:Testcase-Generation}}

There are several examples of model-based test generation tools for Android where a \textit{UI} model is a vital requirement \cite{takala2011experiences,jensen2013automated,banerjee2014detecting,yang2013grey}. AMOGA generates a UI state model in which edges in the model correspond to a test case. To demonstrate the use of the generated model, we have implemented a test generation tool to generate test cases. The tool traverses paths through the model and creates test cases. A test case includes a whole path of edges from an initial to a final state or a sub-path including one origin and one target state. The test cases were implemented using the JUnit format that can be executed with the Robotium testing framework. This is a typical example. For all edge \textit{$e$} = \textit{$e_{1}$, }$e_{2}$,\dots , the event id $e(e_{i})$ is translated to corresponding Robotium API calls that can trigger the event. 

\subsection{Tool Implementation\label{subsec:Tool-Implementation}}

This section provides details of the implementation of the proposed hybrid approach in a tool called Automated Model Generator for Android (\textit{AMOGA}). \textit{AMOGA} was implemented using the Java programming language. It comprises a static analyzer that statically extracts mobile app\textquoteright s supported events which can be supplied as input to the UI crawler. The following paragraphs describe the static analyzer and the UI crawler in details.

\paragraph{Static Analyzer }

To identify the set of events supported by a mobile app, the static analyzer performs callback control flow analysis to generate the WTG \cite{Shengqian2015static} of the application. The static analyzer performs the analysis and builds the graph with the help of GATOR \cite{GATORWebSite,Shengqian2015static}, a Program Analysis Toolkit For Android that we customized to track the events. First, the static analyzer decompiles the apk to extract the bytecode and then starts the analysis. Second, it performs the control flow analysis to generate the CFG and WTG. Last, we applied our events tracking algorithm described in \ref{subsec:Events-Tracking-Algorithm} on the WTG to generate a list of events that includes the UI and system events.

\paragraph{UI Crawler}

The UI crawler implements the mobile app\textquoteright s UI exploration using the crawling algorithm described in \ref{subsec:Crawling-Algorithm}. It is built on top of the Robotium framework\footnote{https://github.com/robotiumtech/robotium/wiki} and utilizes its capabilities of extracting the UI widgets (e.g., checkboxes, buttons, spinners) and fire the actions on their event handlers. We observed that during the crawling process, the number of states might be huge depending on the size of the AUT, with some states appearing more than one. To avoid the state exploration problem, the crawler analyses a new state in (line 8) of the algorithm \ref{subsec:Crawling-Algorithm} to verify whether the state is visited using the visitedSet $vS$ list that is used to store all visited states. This enables the crawler to differentiate a newly discovered state with other explored states before adding to the model and the $vS$ will be updated accordingly.

\section{Evaluation\label{sec:Experimetal-Evaluation}}

This section reports two experimental evaluation that involves code coverage and mutation testing. The first set of the experiment described in subsection \ref{subsec:Study-1:-Measuring-code-coverage} evaluates the model generation capability of our approach and compares it with the state-of-the-art approaches. Specifically, the experiment evaluates the quality of the generated model by measuring the code coverage achieved and the model exploration time for a set of popular apps. The tools selected for the comparison are Monkey \cite{GoogleCodeWebSite}, AndroidGUITAR\cite{SourceforgeGuitarPage}, SwiftHand \cite{choi2013guided}, ORBIT \cite{yang2013grey}, MCrawlT \cite{salva2016model} and MobiGUITAR \cite{Amalfitano2015}. The AndroidGUITAR and ORBIT are not available freely to download. Therefore, their results are taken from the published papers directly without implementation in our environment. The second set of the experiment described in subsection \ref{subsec:Study-2:-Applying-mutation-testing}, involves the application of mutation testing to evaluate the fault detection ability of our approach. Mutation testing is a well-known fault-based testing approach in which numerous form of faults can be induced into the code of an app. A given test suite is applied to test the app to identify the faults induced. Other start-of-the-art approaches performed faults detection by studying the bugs report of an app to identify faults that were previously observed in the app. They introduced these faults into the app and evaluates whether the test suite can detect them. Therefore, we do not intend to make a comparison of the second set of experiments with other approaches/tools.

The experiments were conducted on Linux machine with 64-bit Ubuntu on an i7 Intel processor 2 cores, with 8GB memory and HDD. We used
an emulator configured based on x86\_64 configuration with the Lollipop version (API Level 21) of Android. The experiments intend to answer
the following research questions (RQ): 
\begin{itemize}
\item \textbf{RQ1.} Does our approach produces an informative model that have high coverage of application's behavior?
\item \textbf{RQ2.} What is the fault detection capability of the tests generated by the approach? 
\end{itemize}

\subsection{Study 1: Measuring Code Coverage (RQ1)\label{subsec:Study-1:-Measuring-code-coverage}}

Code coverage has been very useful in accessing the effectiveness of testing approach by many researchers in the literature such as \cite{nguyen2014guitar,Amalfitano2015}. The more the code coverage, the better the potential of a testing approach. A problem that may lead to incorrect behavior due to programming errors may occur in an application. When part of the code is not covered during the testing, such a problem may not be detected. Hence, it is essential to ensure that a test can cover a significant amount of the source code. We used EMMA\footnote{http://emma.sourceforge.net/downloads.html} tool to generate the code coverage. It is an open-source tool that measures and reports Java code coverage for Java applications which is now included in the Android SDK. Emma provides coverage reports (in percentage) at the class, method, basic block, and statement/line levels. We used the statement coverage metric in our experiments for the evaluation of the code coverage results. The statement coverage is used to measure the number of executed statements in source code.

To demonstrate the effectiveness of AMOGA, we conducted experiments on 15 selected mobile apps to measure the achieved code coverage and the model crawling time and compared it with the selected tools for the automated model generation. These apps were used for many experiments in the literature and they are used to evaluate the selected tools. Table \ref{tab:Characteristics-of-applications} presents the characteristics of the selected apps. The table gives a description of the apps in \textit{column} 2, the source line of code (SLOC) in \textit{column} 3, the number of Activities in \textit{column} 4, the category of mobile apps in \textit{column} 5, and the number of downloads based on Google Play analytic as of February 2018 in the last \textit{column}. The selection covers a range of real-world open source mobile apps and falls across different categories such as productivity, education, and finance.

\begin{table*}
\caption{\label{tab:Characteristics-of-applications}Characteristics of the applications used in the experimental evaluation}

\centering{}%
\begin{tabular}{|l|l|c|c|c|c|}
\hline 
Subjects & Description & SLOC & Activities & Category & Download\tabularnewline
\hline 
\hline 
TippyTipper & Tip calculator & 2238 & 5 & Finance  & 100,000+\tabularnewline
\hline 
ToDoManager & Daily task manager & 323 & 2 & Productivity & 100,000+\tabularnewline
\hline 
ContactManager & Contacts manager & 497 & 2 & Tool & 50,000+\tabularnewline
\hline 
Tomdroid & Note taking app & 3711 & 5 & Productivity & 10,000+\tabularnewline
\hline 
AardDict & Dictionary an online Wiki & 4518 & 4 & Books \& Ref. & 10,000+\tabularnewline
\hline 
OpenManager & Simple file manager & 1595 & 6 & Productivity & 5M+\tabularnewline
\hline 
Notepad & Note taking app & 332 & 3 & Productivity & 500,000+\tabularnewline
\hline 
Aagtl  & Navigator & 43105 & 3 & Tool & 10,000+\tabularnewline
\hline 
Explorer & Organizing internal storage and SD card contents & 2844 & 8 & Productivity  & 10M+\tabularnewline
\hline 
Weight  & Weight tracking tool & 2248 & 5 & Health \& fitness & 500,000+\tabularnewline
\hline 
anyMemo & Flash card learning software & 71072 & 8 & Education & 100,000+\tabularnewline
\hline 
MultiSMS & Sending message to multiple users & 1943 & 6 & Communication & 50,000+\tabularnewline
\hline 
Netcounter & Network traffic counter for EDGE/3G and Wi-Fi & 7278 & 3 & Tool & 1M+\tabularnewline
\hline 
MunchLife & Character level tracker for card game & 567 & 2 & Game & 10,000+\tabularnewline
\hline 
AlarmKlock & Alarm clock & 382 & 5 & Tool & 500,000+\tabularnewline
\hline 
\end{tabular}
\end{table*}

\subsubsection{Experimental Results\label{subsec:Experimental-results}}

\begin{figure}
\begin{centering}
\includegraphics[scale=0.35]{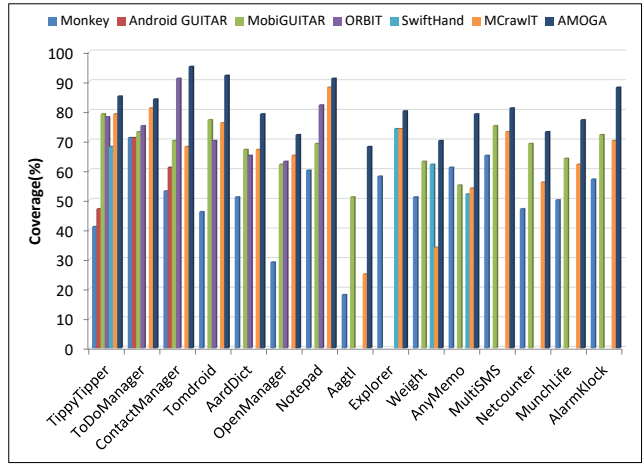}
\par\end{centering}
\caption{\label{fig:Code-coverage}Code coverage of tools}
\end{figure}

Figure \ref{fig:Code-coverage} reports the percentage code coverage obtained by all the selected tools on the applications used for the study. The coverage results show that AMOGA achieved a minimum coverage of 68\% on Aagtl app and a maximum of 95\% on the contactManager app. As we can see from the results, MCrawlT with 88\% and ORBIT with 91\% have a maximum coverage that is close to that of AMOGA, followed by MobiGUITAR with a maximum coverage of 79\% and SwiftHand with 74\%. AndroidGUITAR and Monkey with 71\% each are inferior to other approaches.

\subsubsection{Comparison with state-of-the-art Approaches\label{subsec:Comparison-with-state-of-the-art}}

We statistically compared the coverage results of AMOGA with the state-of-the-art approaches (tools). It is observed that the monkey tool was used as a reference for evaluation in most of the Android testing tools. It can be considered as a baseline because it comes with the Android SDK and it is popular among developers. In the selection, we considered tools whose main goal is to produce a model of an application as an artifact for model-based testing as shown in Table \ref{tab:Comparison-of-features}. A2T2 and A3E are not considered in the comparison, because A2T2 is not available and the coverage result is not published in the literature, while A3E does not produce a model but systematically explore an app to measure the Activity and method coverage respectively.

\begin{figure}
\begin{centering}
\includegraphics[scale=0.3]{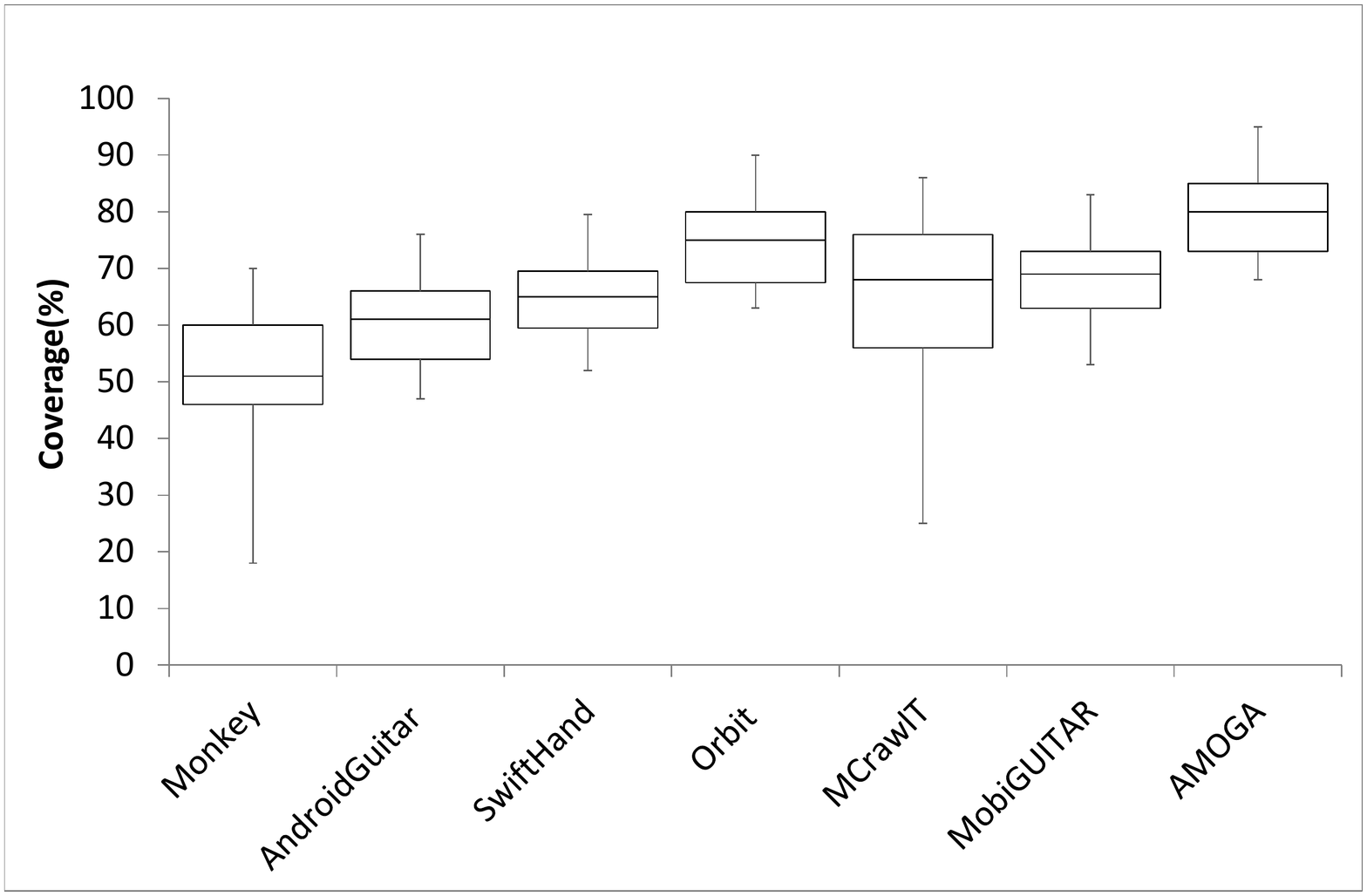}
\par\end{centering}
\caption{\label{fig:Comparison-of-coverage-with-other-tools}Comparison of
coverage of AMOGA with other tools }
\end{figure}

Figure \ref{fig:Comparison-of-coverage-with-other-tools} presents the results of the comparison of AMOGA with the selected tools. The horizontal axis shows the tools used in the comparison. The vertical axis shows the percentage coverage. The boxes give the minimum, average and maximum coverage achieved by the tools. The coverage range achieved by AMOGA for most of the applications is between 73\% and 95\% while MobiGUITAR achieved between 63\% and 73\%, MCrawlT achieved between 56\% and 76\%, ORBIT achieved between 67\% and 80\%, SwiftHand achieved between 59\% and 69\%, AndroidGUITAR achieved between 54\% and 66\%, and Monkey achieved 46\% and 60\%. On average, AMOGA achieved 80\% coverage while ORBIT has 74\%, MobiGUITAR has 69\%, MCrawlT has 68\%, SwiftHand has 65\%, AndroidGUITAR has 61\% and Monkey has achieved 51\% coverage. The results show that the average coverage of ORBIT is close to that of AMOGA. However, based on the range of their coverage, we observed that the lowest coverage obtained by AMOGA which is for Aagtl is higher than the lowest coverage of ORBIT which is on Aarddict. Based on the experimental results in Figures \ref{fig:Code-coverage} and \ref{fig:Comparison-of-coverage-with-other-tools} AMOGA achieved higher coverage than other state-of-the-art tools. In comparison to all the tools, AMOGA stands out as the tool with the highest coverage. This shows that the Static-Dynamic hybridization can enable the generation of more comprehensive models that reflect the behavior of the application.

\begin{figure}
\begin{centering}
\includegraphics[scale=0.3]{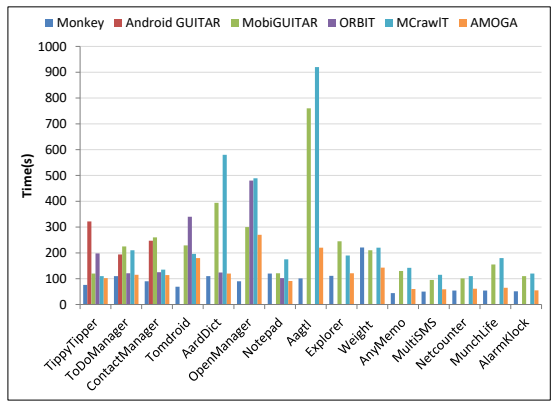}
\par\end{centering}
\caption{\label{fig:Exploration-time-of-tool}Exploration time of tools}
\end{figure}

\begin{figure}
\begin{centering}
\includegraphics[scale=0.3]{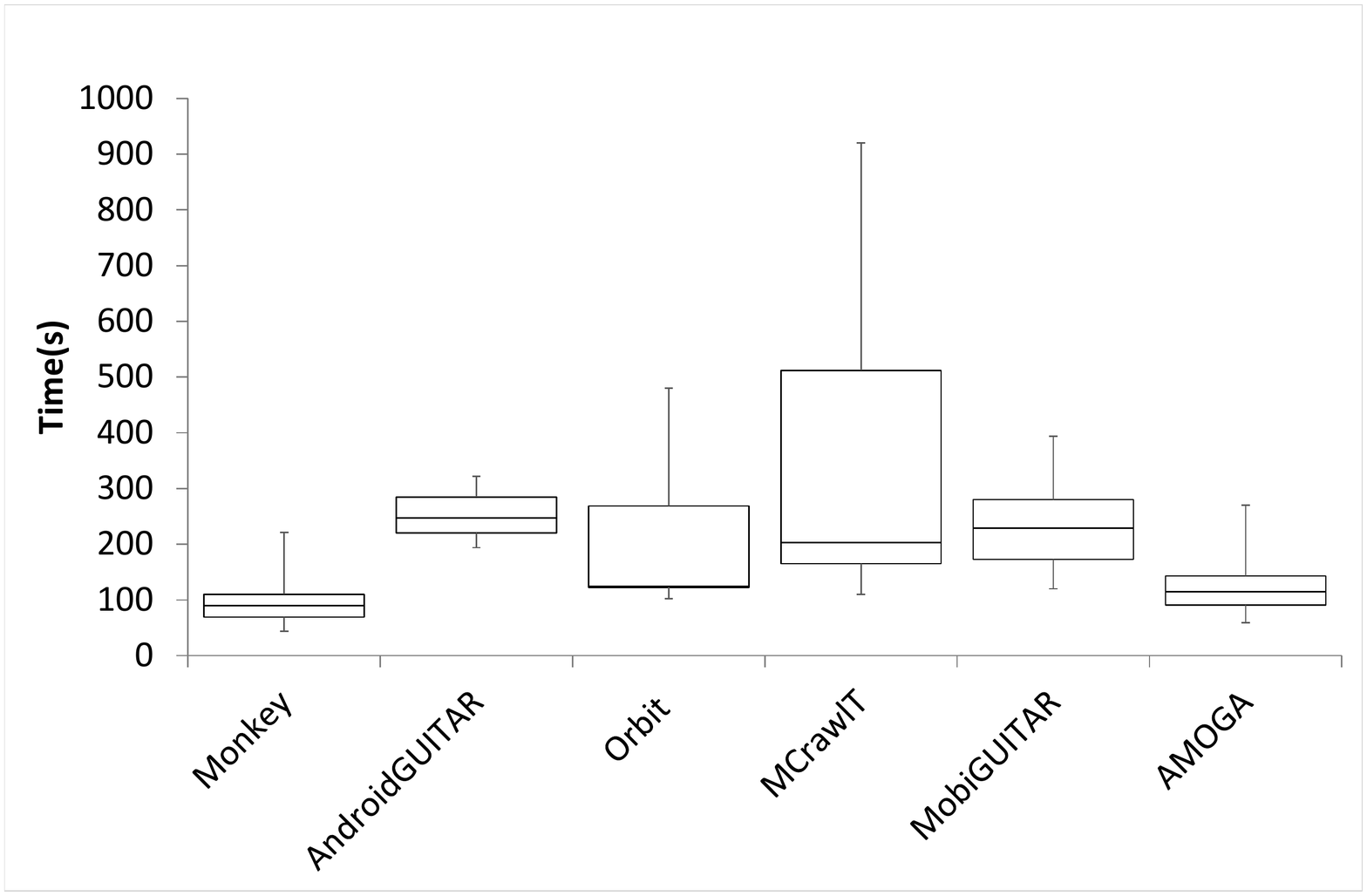}
\par\end{centering}
\caption{\label{fig:Comparison-of-exploration-time-statistics}Comparison of
exploration time of AMOGA with other tools (in seconds) }
\end{figure}

The efficiency of an approach depends on how fast it can explore an application. Figure \ref{fig:Exploration-time-of-tool} reports the exploration time of tools on the 15 applications. AMOGA have the lowest exploration time of 61s for Netcounter app and the maximum is 270s for OpenManager app. The maximum time for Monkey is 221s, AndroidGUITAR have 322s, MobiGUITAR 760s, ORBIT 480s, and MCrawlT have 920s. In addition, the results collected in Figure \ref{fig:Comparison-of-exploration-time-statistics} presents a statistical comparison of the exploration time of AMOGA with other tools. When the results are compared with other tools, Monkey tool explored the applications in an average time of 95s while AMOGA explored the applications in an average time of 127 seconds, ORBIT has 212 seconds, MobiGUITAR is 235 seconds, AndroidGUITAR is 254 seconds and MCrawlT 351 seconds. SwiftHand was designed to run for 3 hours after which it will terminate. This setting enables SwiftHand to run all the applications used in the experiments. When we perform a side by side comparison of AMOGA with other tools, it was observed that the monkey is faster on almost all the applications due to its random nature of the exploration. However, the achieved code coverage is low. Whereas, our strategy performance degrades due to its in-depth exploration process. Nonetheless, AMOGA is almost 2x faster than its hybrid counterpart ORBIT.

\subsection{Study 2: Mutation testing (RQ2)\label{subsec:Study-2:-Applying-mutation-testing} }

Recently, several researchers and practitioners argued about using code coverage alone to validate the quality of a testing tool. It is conceded by many researchers in the literature that, simply testing statements coverage is insufficient to assure an application's quality (to make sure it functions correctly) \cite{deng2017mutation,inozemtseva2014coverage}. Mutation testing represents a complement to the process by ensuring that the apps function as expected and are released without faults. The primary goal of mutation testing is to determine the fault detection ability of a give tool/test suite. Any test that kills the generated mutants is expected to expose many faults in an app \cite{deng2017mutation}. The effectiveness of mutation testing is evaluated by measuring the mutation score (MS) achieved. A mutation score of MS = 1.000 indicates that all the mutants in an app are killed. 

Here, we used mutation testing to evaluate the fault detection capability of our approach to answer RQ2. The study involves the generation of mutants for the 15 selected applications in study 1 to test them and measure the mutation score (MS). Several mutants generation methods/operators were defined and developed for mutation testing such as muJava \cite{ma2006mujava}, the technique by Nilsson and Offutt \cite{nilsson2007automated} that ware designed for desktop applications. However, these techniques are not suitable for mobile apps. The respective counterparts of such tools for mobile apps have been proposed, such as AndroidMutants \cite{deng2017mutation} and muDroid \cite{weiMudroid} which are based on the same principles (mutation operators that are responsible for altering the code of an application) as for desktop applications.

In this study, we used the available tool, muDroid \cite{weiMudroid} to generate mutants and run tests on the selected applications. In muDroid six (6) mutation operators are defined and implemented for the Android apps that work at Smali code (a code created by decompiling .DEX) level. The operators are ICR (Inline Constant Replacement), NOI (Negative Operator Inversion), LCR (Logical Connector Replacement), AOR (Arithmetic Operator Replacement), ROR (Relational Operator Replacement) and RVR (Return Value Replacement). ICR changes the value of a constant before it gets assigned to a variable, NOI inverts a negation of variables, LCR replaces the logical connectors from one to another, AOR replaces the arithmetic operators, ROR replaces the relational operator from one to another and RVR replace the return values to 0 or null.

By default, muDroid generates hundreds of mutants for an application, but it employs a selection criterion for selecting the mutants to reduce the total number of mutants to be used in the testing step. The selection criterion tries to find a small set of mutation operators with which no significant loss of test effectiveness will be observed by selecting representative mutants across the different mutation operators. Table \ref{tab:Mutants-generated-for} shows the results of mutants generation. \textit{Column }2 gives the number of total mutants generated. \textit{Column} 3 gives the number of mutants generated for each operator.

\begin{table*}
\caption{\label{tab:Mutants-generated-for}Mutants generated for the apps }

\centering{}%
\begin{tabular}{|c|c|c|c|c|c|c|c|}
\hline 
\multirow{2}{*}{App} & \multirow{2}{*}{Total Mutants} & \multicolumn{6}{c|}{Mutants for each mutation operators}\tabularnewline
\cline{3-8} \cline{4-8} \cline{5-8} \cline{6-8} \cline{7-8} \cline{8-8} 
 &  & ICR & NOI & LCR & AOR & ROR & RVR\tabularnewline
\hline 
\hline 
TippyTipper & 980 & 113 & 0 & 8 & 83 & 109 & 57\tabularnewline
\hline 
ToDoManager & 371 & 111 & 0 & 1 & 0 & 51 & 24\tabularnewline
\hline 
ContactManager & 156 & 27 & 3 & 5 & 14 & 11 & 29\tabularnewline
\hline 
Tomdroid & 1557 & 257 & 0 & 34 & 102 & 111 & 201\tabularnewline
\hline 
AardDict & 1834 & 367 & 0 & 22 & 39 & 279 & 84\tabularnewline
\hline 
OpenManager & 2423 & 375 & 0 & 46 & 70 & 383 & 77\tabularnewline
\hline 
Notepad & 3286 & 688 & 0 & 40 & 61 & 492 & 109\tabularnewline
\hline 
Aagtl  & 5225 & 658 & 59 & 105 & 325 & 865 & 339\tabularnewline
\hline 
Explorer & 1637 & 308 & 0 & 8  & 65 & 232 & 136\tabularnewline
\hline 
Weight  & 1849 & 296 & 4 & 8 & 209 & 183 & 40\tabularnewline
\hline 
anyMemo & 10850 & 192 & 0 & 11 & 110 & 668 & 105\tabularnewline
\hline 
MultiSMS & 650 & 194 & 0 & 1 & 17 & 91 & 32\tabularnewline
\hline 
Netcounter & 1814 & 433 & 7 & 17 & 43 & 239 & 140\tabularnewline
\hline 
MunchLife & 187 & 19 & 0 & 3 & 10 & 12 & 4\tabularnewline
\hline 
AlarmKlock & 2449 & 89 & 1 & 18 & 37 & 46 & 121\tabularnewline
\hline 
Total  & 35268 & 4127 & 74 & 327 & 1185 & 3772 & 1498\tabularnewline
\hline 
\end{tabular}
\end{table*}

\begin{table*}
\caption{\label{tab:Results-of-mutation}Results of mutation testing}

\centering{}%
\begin{tabular}{|c|c|c|c|c|c|c|c|c|c|}
\hline 
\multirow{2}{*}{App} & \multicolumn{2}{c|}{Mutants} & \multicolumn{6}{c|}{Mutation score of operators} & \multirow{2}{*}{Final MS}\tabularnewline
\cline{2-9} \cline{3-9} \cline{4-9} \cline{5-9} \cline{6-9} \cline{7-9} \cline{8-9} \cline{9-9} 
 & Selected  & Killed  & ICR & NOI & LCR & AOR & ROR & RVR & \tabularnewline
\hline 
\hline 
TippyTipper & 370 & 370 & 0.305 & 0.000 & 0.022 & 0.224 & 0.295 & 0.154 & 1\tabularnewline
\hline 
ToDoManager & 187 & 187 & 0.594 & 0.000 & 0.005 & 0.000 & 0.273 & 0.128 & 1\tabularnewline
\hline 
ContactManager & 89 & 89 & 0.304 & 0.036 & 0.052 & 0.163 & 0.120 & 0.326 & 1\tabularnewline
\hline 
Tomdroid & 705 & 705 & 0.364 & 0.000 & 0.048 & 0.145 & 0.158 & 0.285 & 1\tabularnewline
\hline 
AardDict & 791 & 791 & 0.464 & 0.000 & 0.028 & 0.049 & 0.352 & 0.106 & 1\tabularnewline
\hline 
OpenManager & 951 & 951 & 0.394 & 0.000 & 0.048 & 0.074 & 0.403 & 0.081 & 1\tabularnewline
\hline 
Notepad & 1390 & 1390 & 0.495 & 0.000 & 0.029 & 0.044 & 0.354 & 0.078 & 1\tabularnewline
\hline 
Aagtl  & 2351 & 2351 & 0.280 & 0.025 & 0.045 & 0.138 & 0.368 & 0.144 & 1\tabularnewline
\hline 
Explorer & 749 & 749 & 0.411 & 0.000 & 0.011 & 0.087 & 0.310 & 0.182 & 1\tabularnewline
\hline 
Weight  & 740 & 740 & 0.400 & 0.005 & 0.011 & 0.282 & 0.247 & 0.054 & 1\tabularnewline
\hline 
anyMemo & 1086 & 1086 & 0.177 & 0.000 & 0.010 & 0.101 & 0.615 & 0.097 & 1\tabularnewline
\hline 
MultiSMS & 335 & 290 & 0.433 & 0.000 & 0.000 & 0.025 & 0.210 & 0.108 & 0.866\tabularnewline
\hline 
Netcounter & 879 & 879 & 0.492 & 0.008 & 0.019 & 0.049 & 0.271 & 0.159 & 1\tabularnewline
\hline 
MunchLife & 48 & 48 & 0.395 & 0.000 & 0.062 & 0.208 & 0.250 & 0.083 & 1\tabularnewline
\hline 
AlarmKlock & 312 & 312 & 0.285 & 0.003 & 0.057 & 0.118 & 0.147 & 0.387 & 1\tabularnewline
\hline 
Total  & 10983 & 10938 & \multicolumn{6}{r|}{Average MS} & 0.90\tabularnewline
\hline 
\end{tabular}
\end{table*}

Table \ref{tab:Results-of-mutation} presents the results of the mutation testing. \textit{Column} 2 shows the number of mutants selected and killed. \textit{Column} 3 shows the mutation score obtained for each operator. \textit{Column} 4 presents the total MS obtained for each application. Based on the results presented in Table \ref{tab:Results-of-mutation}, an MS= 1.000 is obtained for all the applications except MultiSMS, meaning that all mutants injected in the applications have been killed. MultiSMS has the lower MS= 0.866 which means that some of the mutants injected in the application were not killed. Out of the 10578 mutants selected for the mutation testing, 45 were not killed which are from the MultiSMS app. For this viewpoint, we can conclude that our approach, AMOGA has yielded high mutant coverage for all the applications except MultiSMS. Therefore it can reveal many faults in mobile apps.

\subsection{Discussion\label{subsec:Discussion}}

The results in Figure \ref{fig:Code-coverage} indicated that AMOGA achieved a maximum coverage of 95\%. This is attributed to our static analysis that generates events to support the exploration and the crawling algorithm for the model exploration. Although ORBIT can achieve a reasonable coverage on some applications with an average close to that of AMOGA, its coverage is low for some other applications due to its inability to analyze other events such as systems events (e.g., lifecycle events). Other tools that dynamically analyze an application to generate inputs for the crawler that is responsible for exploring the events have other limitations. Those tools are not able to explore many events because the availability of some events depends on the existence of other events (as discussed in section \ref{sec:Background}). In our proposed approach, the input events to explore an application are acquired statically from the bytecode. To efficiently explore an app UI taking into account the order of event sequence, AMOGA implements Dijkstra's search algorithm which is an improved form of BFS that uses a priority queue to select the next item to explore. 

From Figure \ref{fig:Code-coverage}, we can see that AMOGA achieved the lowest coverage of 68\%, and 79\% for Aagtl and AnyMemo apps respectively. We manually examine and inspect the applications and we found that Aagtl requires user input for some events to run to completion. AMOGA does not generate concrete user inputs. Hence, the execution paths that require user input may never be covered. AnyMemo is a complex app that allows a user to access many repositories to download dictionaries and wordlists. It relies on the dynamic manipulation of network connectivity which causes difficulty in reaching parts of the application.

Similarly, for the AardDict, Aagtl, multiSMS that process both UI and system events such as life-cycle event due to interrupt and Android broadcast messages, ORBIT, MCrawlT, and MobiGUITAR obtains low coverage, because they do not have support for system events. Monkey offers support for system events; however, it supports a limited number of system events. In contrast, AMOGA supports these events. Hence, it offers better coverage of these applications. AMOGA stands out as the tool with the highest coverage.

Based on the results in Figures \ref{fig:Comparison-of-coverage-with-other-tools} and \ref{fig:Comparison-of-exploration-time-statistics}, we can observe that, on average, AMOGA achieves higher code coverage within a shorter time compared to other tools. On average, AMOGA achieves 80\% coverage within an average time of 127 seconds, whereas ORBIT, MobiGUITAR, MCrawlT, and AndroidGUITAR achieve 74\%, 68\%, 64\%, and 59\% respectively, within an average time 212 seconds, 235 seconds, 351 seconds and 254 seconds. When we compare the time taken by every application involved in our experiment, we can see that the minimum time is 59 seconds for the MultiSMS app and the maximum time is 270 seconds for the OpenManager app. OpenManager takes longer time due to its large number of Activities that have complex UI. Thus, the time along with the coverage obtained indicates that AMOGA is better than the other tools in both coverage and execution time. 

The result of the mutation testing in Table \ref{tab:Results-of-mutation} shows that 45 mutants in MultiSMS app were not killed. After inspecting the mutants generated for the app, we detected that the mutants are equivalent mutants, i.e., semantically equivalent to the original code. 

The intent objects that are used to specify the target activity have two primary forms; the explicit intent which is used for intra-app interactions, i.e., when the target component is inside the same app and the implicit intent that is used for inter-app communication. The current implementation of AMOGA handles only intra-app events which are processed via the explicit intents.

\subsection{Threats to Validity}

The section discusses the threats to the validity of the study related to the evaluation reported in two studies. The study presents threats that are discussed as follows. First, in respect to the generated test, some of the events may need additional input from the tester to run. An example is an options window that may require a tester to select an item from the options. To ensure they run as expected, we implemented an automated user input generation system that generates input randomly. Second, is the selection of subjects (mobile apps) used to run the experiments which may affect the results. As with most software engineering research, it is difficult to ensure the representativeness of the selected subjects. We have selected mobile apps of different sizes, from a different category, that were used to validate several tools by previous researchers. This can provide consistency across multiple studies.

Finally, we have considered six state-of-the-art tools (e.g., Monkey, AndroidGUITAR, SwiftHand, ORBIT, MCrawlT, and MobiGUITAR) using a set of common case studies with the same code coverage reports to indicate the performance of AMOGA. Although may be relevant to our work, we have not considered applying the common case studies using A3E owing to differences in the coverage reports (i.e., based on method and activity). For this reason, a fair comparison of AMOGA with A3E may be difficult (e.g., method coverage does not imply statement coverage and vice versa). To do so, we may need to enhance A3E to incorporate similar code coverage, which is a difficult endeavor.

\section{Related Work\label{sec:Related-Work}}

We summarized the related works on the MBT of Mobile apps based on two categories: the dynamic approaches and hybrid approaches. Due to the weaknesses of dynamic approaches, the hybrid approaches have drawn much attraction from researchers in recent years.

\subsection{Dynamic Approaches\label{subsec:Dynamic-Approaches}}

One of the earliest techniques is the GUI ripping by Memon \textit{et al.} \cite{memon2003gui} that was implemented as part of GUITAR tool to automate GUI exploration for desktop applications \cite{nguyen2014guitar}. GUITAR consists of a ripper that generate event-flow graph (EFG) model by automatically interacting with an application during execution to extract all relevant information about its GUI which is later converted into test cases. An extension of the tool has been proposed for the Android platform known as AndroidGUITAR \cite{nguyen2014guitar,SourceforgeGuitarPage}. While the tool provides suitable exploration strategies for GUI applications with traditional GUI design for desktop applications, some factors pose difficulties when using it to explore touch-based smartphone applications. Smartphone apps consist of a rich set of user input features such as gestures (e.g., swipes, pinches) that are not firmly confined to a particular GUI object (e.g., text box or button). Furthermore, the UI state can be changed by another application's component in Android app or by a service running in the background. This can be handled through the system-wide callback by creating a list of action sequences that can be executed by a user on a UI. In view of this, only a subset of the UI states can be explored because the behavior of the callback requests can certainly modify UI states \cite{Yan2015static}. Hence, this approach is not able to capture the rich set of user inputs associated with a mobile app. In addition, it is difficult to explore the infeasible paths because the visibility of some elements depends on other elements in the UI. For example, some modal UI (dialog box) can sometimes be visible or invisible. Toggling their visibility can expose or hide some events. The static analysis employed in the proposed approach has dealt with this issue. 

Android Automatic Testing Tool (A2T2) by Amalfitano \textit{et al.} \cite{amalfitano2011gui} is based on the dynamic analysis of an application at the run-time. It uses a crawler that simulates real events of the user on the UI to generate test cases that can be automatically executed on an application for crash testing and regression testing. It generates a GUI tree, a model that can be used for driving test cases automatically. The current implementation of A2T2 focused on the user events triggered through the GUI only but does not consider other types of events supported by a mobile app such as inter-component communication within the application, external events invoked by the hardware sensors or network. Therefore, it does not have support for the rich set of inputs associated with
Android apps. 

AndroidRipper \cite{amalfitano2012using} dynamically analyzes an application using a ripper that systematically rips the application\textquoteright s UI to generate test cases for stress-testing that can reveal unexpected faults in the code. The goal of the AndroidRipper is to generate and execute tests automatically but does not develop a reusable model of an application. Amalfitano \textit{et al.} proposed MobiGUITAR \cite{Amalfitano2015}, an extension of AndroidRipper that implements an algorithm based on breadth-first traversal to dynamically traverse an application to create a tasks list consisting of sequences of events. The tasks list is used to fire events on the UI to generate a state model of the application that can be used for test case generation. 

Swifthand \cite{choi2013guided} tool dynamically analyzes a given application to generate sequences of test inputs for Android apps. It implemented an algorithm that is based on machine learning to learn an approximate model of the application during testing. The learned model is used to generate event sequences that can visit the unexplored
states and execute the generated sequences on the application to refine the model. The authors do not focus on the quality of the generated model but rather to guide the test execution. MCrawlT \cite{salva2016model} learns a model of the navigational paths of mobile apps at run-time. It implemented an algorithm which dynamically analyzes an application to create a tasks list corresponding to the UI of the application. The tasks list is executed on the application using the Robotium framework to explore the UI and infers a model of the application that captures the supported events. The tool only generates default touching and scrolling events from the UI which can be supported by Robotium but does not consider system events. 

Baek and Bae \cite{baek2016automated} proposed an automated testing framework for Android apps utilizing dynamic analysis. They implemented a set of multi-level GUI Comparison Criteria (GUICC) that provides the selection of multiple abstraction levels to improve GUI model generation. QUANTUM \cite{zaeem2014automated} is a framework that receives a model of a mobile app and used it as input for test suite generation that includes oracles for testing the user-interaction features of the application. All the reviewed tools utilized the dynamic analysis approach for the model reverse engineering. Although they have struggled to improve the effectiveness of the UI model, they still suffer from the limitations of dynamic analysis as discussed in \ref{sec:Background}.

Other state-of-the-art tools \cite{machiry2013dynodroid,hao2014puma} are designed to test an application using the dynamic analysis automatically but they do not generate a re-usable model of the application. Dynodroid \cite{machiry2013dynodroid} is based on a random exploration similar to Monkey with more improvement to generate the UI and system events and checks which ones are relevant to the application. Similarly, PUMA \cite{hao2014puma} is a framework that includes a generic UI automator that provides a random exploration of mobile apps using the basic monkey exploration strategy. G{\footnotesize{}VT} \cite{moran2018automated} is an automated approach for verifying the UI of a mobile app against the intended design specifications.

\subsection{Hybrid Approaches\label{subsec:Hybrid-Approaches}}

Considering the hybrid approach, ORBIT tool \cite{yang2013grey} integrated both static and dynamic analysis to generate a state model from Android apps. It leverages the static analysis in WALA framework \cite{WALAframeworkWebSite} to analyze the source code of an application to generate a call graph that is used to generate a set of user actions that are supported by an application. It performs a backward slice on the listener objects to track the view ids of the UIs associated with the listeners. A dynamic crawler (built on top of Robotium) is used to fire actions on the UI objects to explore the application. This generates a state model that can be used for generating the test cases. Although the tool employs this approach, its static analysis missed some vital information because it does not account for the life-cycle events triggered by the life-cycle callback methods. For example, the onCreate method that manages the Activity life-cycle. The Life-cycle callbacks for activities, dialog, and menus can outline major changes to the visible state and behavior of an application.

Azim and Neamtiu \cite{azim2013targeted} proposed A3E (Automatic Android App Explorer) tool that is also based on hybrid static and dynamic analysis to automatically explore a mobile app running on a real phone or an emulator. A3E uses the static analysis in ScanDroid \cite{fuchs2009scandroid,ScanDroidWebSite} to perform data flow analysis (taint tracking) on the bytecode of an application to construct static activity transition graph (SATG) with nodes representing activities and edges showing the possible transitions between the activities. The SATG is then used as input for the systematic exploration of the application to be tested. The automatic explorer rips an application using Troyd tool \cite{TroydWebSite} (which is based on Robotium) to extract GUI elements that are used to fire events on an application. However, as the analysis focuses on applications' Activities, this graph representation does not capture the menus/dialog and does not account for the general UI effects of event handlers such as window-close and the triggered callbacks. Hence, it is less comprehensive as it does not cover some highly-required information. 

\subsection{Comparison of Features of Tools\label{subsec:Comparison-of-Features}}

Table \ref{tab:Comparison-of-features} reports the comparison of the features of the testing tools. \textit{Column 2} shows the type of approach used by a tool. Most of the tools are based on a dynamic approach that analyzes an app at run-time (Black-box). However, ORBIT and A3E used a static-dynamic approach (Gray-box) which require the source code or bytecode in the case of A3E.\textit{ Column 3} indicates the types of events supported by the tool. All the tools support only UI events except Monkey and Dynodroid that provide support for a limited number of system events as they utilized the Random technique that is part of the Android framework. \textit{Column 4 }shows whether the tool produces an artifact or not. Monkey produces only logcat reports, Dynodroid produces logcat reports and EMMA coverage reports and A3E produces Activity coverage and method coverage reports. A2T2 produces GUI Tree, AndroidGUITAR produces Event Flow Graph (EFG) model, MCrawlT produces Parameterised Labeled Transition System (PLTS) model,
while MobiGUITAR, SwinftHand, and ORBIT generates Finite State Machine (FSM).

Table \ref{tab:Compariosn-of-input} gives a comparison of input generation of the testing tools. \textit{Column 2 }shows the input generation method used to supply input for the exploration. All the tools dynamically rip the information and used it as an input for the exploration except ORBIT and A3E which identify the input statically with the help of static analysis tool. ORBIT performs the analysis with the help of WALA framework, A3E uses ScanDroid while our strategy performs the analysis with the help of GATOR. The static analysis we employed handles system events through the analysis of callbacks. \textit{Column 3} states the exploration strategy used to explore an application. Monkey and Dynodroid applied the random exploration strategy while the other tools are based on systematic exploration strategy. \textit{Column 4} indicates the underlying framework on which the exploration by the tool is built. A3E uses Troyd tool and RERAN while ORBIT and AMOGA uses Robotium framework.

Table \ref{tab:Comparison-of-static-dynamic} presents the comparison of the static-dynamic characteristics of the tools in the literature. \textit{Column 2} indicates the input required for the static analysis. ORBIT requires the source code of an application while A3E and AMOGA uses the bytecode of an application. \textit{Column 3} shows which application's component is used for the analysis. ORBIT utilizes an action detector that targets all the entry points of an application. A3E analyze the Activities of an application while AMOGA uses a static analyzer that analyzes all windows of the application (including dialog and menus). \textit{Column 4} shows the output of the static analysis. ORBIT produced a call graph that is used to identify supported actions. A3E generates Static Activity Transition Graph (SATG) while AMOGA generates Windows Transition Graph (WTG). In comparison to the SATG in \cite{azim2013targeted}, the WTG in our approach is derived from the analysis of the key aspects of the UI behavior such as widgets, event handlers, callback sequences, and window stack changes. It captures menus/dialog and model the window stack and its state changes. Our static analysis is more advanced because it takes care of menus/dialog and the system events. Hence, it is more comprehensive as it covers more important information. \textit{Column 5} shows the crawling algorithm used by the tools that utilizes the hybrid approach. ORBIT implemented a modified DFS for crawling to build a model. A3E uses the standard DFS to explore an application. To improve the weakness of the DFS algorithm as discussed in section \ref{subsec:Crawling-Algorithm}, AMOGA uses Dijkstra's algorithm for the crawling.

Given the summary of those tools and approaches in the literature, our approach, AMOGA derived benefits of all the features discussed here. We have included our strategy also in the comparison in the Tables \ref{tab:Comparison-of-features}, \ref{tab:Compariosn-of-input} and \ref{tab:Comparison-of-static-dynamic}. In fact, our approach is different from others in many aspects such as the static analysis used (the component target and the generated output and the crawling approach. For example, some of the recent tools such as MobiGUITAR and MCrawlT are based on dynamic analysis while AMOGA uses static-dynamic analysis. As discussed in section \ref{sec:Background}, the most challenging issue with any UI dynamic analysis technique is the way and order in which UI events are found and fired. The scalability issue that is associated with large the amount of data collected at run time is also challenging, because most of the data generated by dynamic analysis are inaccurate and may need to be weeded out \cite{choudhary2015automated,silva2013combining,kull2012automatic}. The static analysis we employed analyzes the callback methods (both event handler and life-cycle callbacks).

\begin{table*}
\caption{\label{tab:Comparison-of-features}Comparison of features of testing
tools}

\centering{}%
\begin{tabular}{|c|c|c|c|c|c|c|}
\hline 
\multirow{2}{*}{Name} & \multicolumn{2}{c|}{Approach} & \multicolumn{2}{c|}{Events} & \multicolumn{2}{c|}{Produced artifacts}\tabularnewline
\cline{2-7} \cline{3-7} \cline{4-7} \cline{5-7} \cline{6-7} \cline{7-7} 
 & Dynamic & Hybrid & UI & System & Formal model & Others\tabularnewline
\hline 
\hline 
Monkey & \Checkmark{} & x & \Checkmark{} & \Checkmark{} & x & LogCat reports\tabularnewline
\hline 
AndroidGUITAR & \Checkmark{} & x & \Checkmark{} & x & EFG Model & x\tabularnewline
\hline 
A2T2 & \Checkmark{} & x & \Checkmark{} & x & x & GUI Tree\tabularnewline
\hline 
Dynodroid & \Checkmark{} & x & \Checkmark{} & \Checkmark{} & x & LogCat reports, Code coverge\tabularnewline
\hline 
MobiGUITAR & \Checkmark{} & x & \Checkmark{} & x & FSM Model & x\tabularnewline
\hline 
SwiftHand & \Checkmark{} & x & \Checkmark{} & x & FSM Model & x\tabularnewline
\hline 
MCrawlT & \Checkmark{} & x & \Checkmark{} & x & PLTS Model & x\tabularnewline
\hline 
A3E & x & \Checkmark{} & \Checkmark{} & \Checkmark{} & x & Coverage reports\tabularnewline
\hline 
ORBIT & x & \Checkmark{} & \Checkmark{} & x & FSM Model & x\tabularnewline
\hline 
AMOGA & x & \Checkmark{} & \Checkmark{} & \Checkmark{} & FSM Model & x\tabularnewline
\hline 
\end{tabular}
\end{table*}

\begin{table*}
\caption{\label{tab:Compariosn-of-input}Comparison of input generation of
testing tools}

\centering{}%
\begin{tabular}{|c|c|c|c|c|c|}
\hline 
\multirow{2}{*}{Name} & \multicolumn{2}{c|}{Input generation} & \multicolumn{2}{c|}{Exploration strategy} & \multirow{2}{*}{Exploration framework}\tabularnewline
\cline{2-5} \cline{3-5} \cline{4-5} \cline{5-5} 
 & Dynamically & Static analysis & Random & Systematic & \tabularnewline
\hline 
\hline 
Monkey & \Checkmark{} & x & \Checkmark{} & x & Monkey UI exerciser\tabularnewline
\hline 
AndroidGUITAR & \Checkmark{} & x & x & \Checkmark{} & GUIRipper\tabularnewline
\hline 
A2T2 & \Checkmark{} & x & x & \Checkmark{} & GUIRipper\tabularnewline
\hline 
Dynodroid & \Checkmark{} & x & \Checkmark{} & x & MonkeyRunner\tabularnewline
\hline 
MobiGUITAR & \Checkmark{} & x & x & \Checkmark{} & AndroidRipper\tabularnewline
\hline 
SwiftHand & \Checkmark{} & x & x & \Checkmark{} & Chipmpchat\tabularnewline
\hline 
MCrawlT & \Checkmark{} & x & x & \Checkmark{} & Robotium\tabularnewline
\hline 
A3E & x & Scandroid & x & \Checkmark{} & Troyd + RERAN\tabularnewline
\hline 
ORBIT & x & WALA framework & x & \Checkmark{} & Robotium\tabularnewline
\hline 
AMOGA & x & GATOR & x & \Checkmark{} & Robotium\tabularnewline
\hline 
\end{tabular}
\end{table*}

\begin{table*}
\caption{\label{tab:Comparison-of-static-dynamic}Comparison of the characteristics
of static-dynamic analysis of tools}

\centering{}%
\begin{tabular}{|c|c|c|c|c|}
\hline 
Name & Input for static analysis & Target & Output of static analysis & Crawling algorithm\tabularnewline
\hline 
\hline 
ORBIT & Source code & Entry points & Call graph & DFS\tabularnewline
\hline 
A3E & Bytecode & Activities & SATG & DFS\tabularnewline
\hline 
AMOGA & Bytecode & Windows & WTG & Dijkstra\tabularnewline
\hline 
\end{tabular}
\end{table*}

\section{Conclusion\label{sec:Conclusion}}

In this paper, we presented AMOGA, a tool that is based on a hybrid static-dynamic approach for automated UI model generation from mobile apps. AMOGA consists of a static analyzer that is responsible for extracting events supported by a mobile app and a dynamic crawler that systematically crawls and generates a model of the application. We applied the tool to 15 real-world mobile apps to generate a model of the UI. Results from the experimental evaluation showed that AMOGA can generate a comprehensive model that can improve the code coverage of an app coverage compared to other state-of-the-art tools. We also performed mutation testing to identify the fault detection ability of our approach. The results indicated that AMOGA achieves a high mutation score which indicates it can reveal many faults in an app.

For future work, we intend to extend AMOGA to handle inter-app events, to/from other applications which are processed via the implicit intents. This will expand the number of currently supported system events which will enable AMOGA to explore apps that support inter-app communication as well. Consequently, it will expand the range of mobile apps that could be explored by our tool and further increase the code coverage.

\section*{Acknowledgments}

The work reported in this paper is funded by Fundamental Research Grant from the Ministry of Higher Education Malaysia titled: A Reinforcement Learning Sine Cosine based Strategy for Combinatorial Test Suite Generation. We thank MOHE for the contribution and support, Grant number: RDU170103.

\bibliographystyle{IEEEtran}
\bibliography{myBibTest}

\begin{thebibliography}{10}
\providecommand{\url}[1]{#1}
\csname url@samestyle\endcsname
\providecommand{\newblock}{\relax}
\providecommand{\bibinfo}[2]{#2}
\providecommand{\BIBentrySTDinterwordspacing}{\spaceskip=0pt\relax}
\providecommand{\BIBentryALTinterwordstretchfactor}{4}
\providecommand{\BIBentryALTinterwordspacing}{\spaceskip=\fontdimen2\font plus
\BIBentryALTinterwordstretchfactor\fontdimen3\font minus
  \fontdimen4\font\relax}
\providecommand{\BIBforeignlanguage}[2]{{%
\expandafter\ifx\csname l@#1\endcsname\relax
\typeout{** WARNING: IEEEtran.bst: No hyphenation pattern has been}%
\typeout{** loaded for the language `#1'. Using the pattern for}%
\typeout{** the default language instead.}%
\else
\language=\csname l@#1\endcsname
\fi
#2}}
\providecommand{\BIBdecl}{\relax}
\BIBdecl

\bibitem{Gartner}
\BIBentryALTinterwordspacing
Worldwide smartphone sales. [Online; accessed 20-June-2018]. [Online].
  Available: \url{https://www.gartner.com/newsroom/id/3876865}
\BIBentrySTDinterwordspacing

\bibitem{GartnerWebSite}
\BIBentryALTinterwordspacing
Worldwide pc sales. [Online; accessed 20-June-2018]. [Online]. Available:
  \url{https://www.gartner.com/newsroom/id/3871149}
\BIBentrySTDinterwordspacing

\bibitem{nayebi2012state}
F.~Nayebi, J.-M. Desharnais, and A.~Abran, ``The state of the art of mobile
  application usability evaluation,'' in \emph{Electrical \& Computer
  Engineering (CCECE), 2012 25th IEEE Canadian Conference on}.\hskip 1em plus
  0.5em minus 0.4em\relax IEEE, 2012, pp. 1--4.

\bibitem{TechCrunchWebSite}
\BIBentryALTinterwordspacing
Worldwide app revenue sales. [Online; accessed 20-June-2018]. [Online].
  Available:
  \url{https://techcrunch.com/2018/01/17/global-app-downloads-topped-175-billion-in-2017-revenue-surpassed-86-billion/}
\BIBentrySTDinterwordspacing

\bibitem{StatistaWebSite}
\BIBentryALTinterwordspacing
Worldwide app revenue forecast. [Online; accessed 20-June-2018]. [Online].
  Available:
  \url{https://www.statista.com/statistics/269025/worldwide-mobile-app-revenue-forecast/}
\BIBentrySTDinterwordspacing

\bibitem{yang2013grey}
W.~Yang, M.~R. Prasad, and T.~Xie, ``A grey-box approach for automated
  gui-model generation of mobile applications.'' in \emph{FASE}, vol.~13.\hskip
  1em plus 0.5em minus 0.4em\relax Springer, 2013, pp. 250--265.

\bibitem{Yan2015static}
S.~Yang, D.~Yan, H.~Wu, Y.~Wang, and A.~Rountev, ``Static control-flow analysis
  of user-driven callbacks in android applications,'' in \emph{Software
  Engineering (ICSE), 2015 IEEE/ACM 37th IEEE International Conference on},
  vol.~1.\hskip 1em plus 0.5em minus 0.4em\relax IEEE, 2015, pp. 89--99.

\bibitem{dehlinger2011mobile}
J.~Dehlinger and J.~Dixon, ``Mobile application software engineering:
  Challenges and research directions,'' in \emph{Workshop on Mobile Software
  Engineering}, vol.~2, 2011, pp. 29--32.

\bibitem{janicki2012obstacles}
M.~Janicki, M.~Katara, and T.~P{\"a}{\"a}kk{\"o}nen, ``Obstacles and
  opportunities in deploying model-based gui testing of mobile software: a
  survey,'' \emph{Software Testing, Verification and Reliability}, vol.~22,
  no.~5, pp. 313--341, 2012.

\bibitem{muccini2012software}
H.~Muccini, A.~Di~Francesco, and P.~Esposito, ``Software testing of mobile
  applications: Challenges and future research directions,'' in
  \emph{Proceedings of the 7th International Workshop on Automation of Software
  Test}.\hskip 1em plus 0.5em minus 0.4em\relax IEEE Press, 2012, pp. 29--35.

\bibitem{osterweil1996strategic}
L.~Osterweil, ``Strategic directions in software quality,'' \emph{ACM Computing
  Surveys (CSUR)}, vol.~28, no.~4, pp. 738--750, 1996.

\bibitem{amalfitano2011gui}
D.~Amalfitano, A.~R. Fasolino, and P.~Tramontana, ``A gui crawling-based
  technique for android mobile application testing,'' in \emph{Software
  Testing, Verification and Validation Workshops (ICSTW), 2011 IEEE Fourth
  International Conference on}.\hskip 1em plus 0.5em minus 0.4em\relax IEEE,
  2011, pp. 252--261.

\bibitem{salva2016model}
S.~Salva, P.~Lauren{\c{c}}ot, and S.~R. Zafimiharisoa, ``Model inference of
  mobile applications with dynamic state abstraction,'' in \emph{Software
  Engineering, Artificial Intelligence, Networking and Parallel/Distributed
  Computing 2015}, R.~Lee, Ed.\hskip 1em plus 0.5em minus 0.4em\relax Springer,
  2016, pp. 177--193.

\bibitem{banerjee2013graphical}
I.~Banerjee, B.~Nguyen, V.~Garousi, and A.~Memon, ``Graphical user interface
  (gui) testing: Systematic mapping and repository,'' \emph{Information and
  Software Technology}, vol.~55, no.~10, pp. 1679--1694, 2013.

\bibitem{lu2012automated}
L.~Lu and Y.~Huang, ``Automated gui test case generation,'' in \emph{Computer
  Science \& Service System (CSSS), 2012 International Conference on}.\hskip
  1em plus 0.5em minus 0.4em\relax IEEE, 2012, pp. 582--585.

\bibitem{kull2012automatic}
A.~Kull, ``Automatic gui model generation: State of the art,'' in
  \emph{Software Reliability Engineering Workshops (ISSREW), 2012 IEEE 23rd
  International Symposium on}.\hskip 1em plus 0.5em minus 0.4em\relax IEEE,
  2012, pp. 207--212.

\bibitem{nguyen2014guitar}
B.~N. Nguyen, B.~Robbins, I.~Banerjee, and A.~Memon, ``Guitar: an innovative
  tool for automated testing of gui-driven software,'' \emph{Automated Software
  Engineering}, vol.~21, no.~1, pp. 65--105, 2014.

\bibitem{wasserman2010software}
A.~I. Wasserman, ``Software engineering issues for mobile application
  development,'' in \emph{Proceedings of the FSE/SDP workshop on Future of
  software engineering research}.\hskip 1em plus 0.5em minus 0.4em\relax ACM,
  2010, pp. 397--400.

\bibitem{baek2016automated}
Y.-M. Baek and D.-H. Bae, ``Automated model-based android gui testing using
  multi-level gui comparison criteria,'' in \emph{Proceedings of the 31st
  IEEE/ACM International Conference on Automated Software Engineering}.\hskip
  1em plus 0.5em minus 0.4em\relax ACM, 2016, pp. 238--249.

\bibitem{utting2012taxonomy}
M.~Utting, A.~Pretschner, and B.~Legeard, ``A taxonomy of model-based testing
  approaches,'' \emph{Software Testing, Verification and Reliability}, vol.~22,
  no.~5, pp. 297--312, 2012.

\bibitem{aho2015making}
P.~Aho, M.~Suarez, A.~Memon, and T.~Kanstr{\'e}n, ``Making gui testing
  practical: Bridging the gaps,'' in \emph{Information Technology-New
  Generations (ITNG), 2015 12th International Conference on}.\hskip 1em plus
  0.5em minus 0.4em\relax IEEE, 2015, pp. 439--444.

\bibitem{amalfitano2012using}
D.~Amalfitano, A.~R. Fasolino, P.~Tramontana, S.~De~Carmine, and A.~M. Memon,
  ``Using gui ripping for automated testing of android applications,'' in
  \emph{Proceedings of the 27th IEEE/ACM International Conference on Automated
  Software Engineering}.\hskip 1em plus 0.5em minus 0.4em\relax ACM, 2012, pp.
  258--261.

\bibitem{grilo2010reverse}
A.~M.~P. Grilo, A.~C.~R. Paiva, and J.~P. Faria, ``Reverse engineering of gui
  models for testing,'' in \emph{Information Systems and Technologies (CISTI),
  2010 5th Iberian Conference on}.\hskip 1em plus 0.5em minus 0.4em\relax IEEE,
  2010, pp. 1--6.

\bibitem{SourceforgeGuitarPage}
\BIBentryALTinterwordspacing
Androidguitar. [Online; accessed 6-November-2017]. [Online]. Available:
  \url{https://sourceforge.net/projects/guitar/files/android-guitar/}
\BIBentrySTDinterwordspacing

\bibitem{tao2016building}
C.~Tao and J.~Gao, ``Building a model-based gui test automation system for
  mobile applications,'' \emph{International Journal of Software Engineering
  and Knowledge Engineering}, vol.~26, no. 09n10, pp. 1605--1615, 2016.

\bibitem{morgado2012dynamic}
I.~C. Morgado, A.~Paiva, and J.~Faria, ``Dynamic reverse engineering of
  graphical user interfaces,'' \emph{Int. Journal on Advances in Software},
  vol.~5, no.~3, pp. 224--246, 2012.

\bibitem{silva2013combining}
C.~E. Silva and J.~C. Campos, ``Combining static and dynamic analysis for the
  reverse engineering of web applications,'' in \emph{Proceedings of the 5th
  ACM SIGCHI symposium on Engineering interactive computing systems}.\hskip 1em
  plus 0.5em minus 0.4em\relax ACM, 2013, pp. 107--112.

\bibitem{salihu2016comparative}
I.~A. Salihu and R.~Ibrahim, ``Comparative analysis of gui reverse engineering
  techniques,'' in \emph{Advanced Computer and Communication Engineering
  Technology}.\hskip 1em plus 0.5em minus 0.4em\relax Springer, 2016, pp.
  295--305.

\bibitem{azim2013targeted}
T.~Azim and I.~Neamtiu, ``Targeted and depth-first exploration for systematic
  testing of android apps,'' in \emph{Acm Sigplan Notices}, vol.~48,
  no.~10.\hskip 1em plus 0.5em minus 0.4em\relax ACM, 2013, pp. 641--660.

\bibitem{Shengqian2015static}
S.~Yang, H.~Zhang, H.~Wu, Y.~Wang, D.~Yan, and A.~Rountev, ``Static window
  transition graphs for android (t),'' in \emph{Automated Software Engineering
  (ASE), 2015 30th IEEE/ACM International Conference on}.\hskip 1em plus 0.5em
  minus 0.4em\relax IEEE, 2015, pp. 658--668.

\bibitem{yang2015static}
S.~Yang, \emph{Static analyses of GUI behavior in Android applications}.\hskip
  1em plus 0.5em minus 0.4em\relax The Ohio State University, 2015.

\bibitem{choudhary2015automated}
S.~R. Choudhary, A.~Gorla, and A.~Orso, ``Automated test input generation for
  android: Are we there yet?(e),'' in \emph{Automated Software Engineering
  (ASE), 2015 30th IEEE/ACM International Conference on}.\hskip 1em plus 0.5em
  minus 0.4em\relax IEEE, 2015, pp. 429--440.

\bibitem{JUnitWebSite}
\BIBentryALTinterwordspacing
Junit. [Online; accessed 6-April-2018]. [Online]. Available:
  \url{https://junit.org/junit5/}
\BIBentrySTDinterwordspacing

\bibitem{SeleniumWebSite}
\BIBentryALTinterwordspacing
Selenium. [Online; accessed 6-April-2018]. [Online]. Available:
  \url{https://www.seleniumhq.org/}
\BIBentrySTDinterwordspacing

\bibitem{GoogleCodeWebSite}
\BIBentryALTinterwordspacing
Monkey: Ui application exerciser. [Online; accessed 5-October-2017]. [Online].
  Available:
  \url{http://developer.android.com/guide/developing/tools/monkey.html}
\BIBentrySTDinterwordspacing

\bibitem{takala2011experiences}
T.~Takala, M.~Katara, and J.~Harty, ``Experiences of system-level model-based
  gui testing of an android application,'' in \emph{Software Testing,
  Verification and Validation (ICST), 2011 IEEE Fourth International Conference
  on}.\hskip 1em plus 0.5em minus 0.4em\relax IEEE, 2011, pp. 377--386.

\bibitem{memon2003gui}
A.~M. Memon, I.~Banerjee, and A.~Nagarajan, ``Gui ripping: Reverse engineering
  of graphical user interfaces for testing.'' in \emph{WCRE}, vol.~3, 2003, p.
  260.

\bibitem{paiva2007reverse}
A.~C. Paiva, J.~C. Faria, and P.~M. Mendes, ``Reverse engineered formal models
  for gui testing,'' in \emph{International Workshop on Formal Methods for
  Industrial Critical Systems}.\hskip 1em plus 0.5em minus 0.4em\relax
  Springer, 2007, pp. 218--233.

\bibitem{limaye2009software}
M.~G. Limaye, \emph{Software testing}.\hskip 1em plus 0.5em minus 0.4em\relax
  Tata McGraw-Hill Education, 2009.

\bibitem{lin2014accuracy}
Y.-D. Lin, J.~F. Rojas, E.~T.-H. Chu, and Y.-C. Lai, ``On the accuracy,
  efficiency, and reusability of automated test oracles for android devices,''
  \emph{IEEE Transactions on Software Engineering}, vol.~40, no.~10, pp.
  957--970, 2014.

\bibitem{liu2009covering}
H.~Liu and H.~B.~K. Tan, ``Covering code behavior on input validation in
  functional testing,'' \emph{Information and Software Technology}, vol.~51,
  no.~2, pp. 546--553, 2009.

\bibitem{aho2014murphy}
P.~Aho, M.~Suarez, T.~Kanstr{\'e}n, and A.~M. Memon, ``Murphy tools: Utilizing
  extracted gui models for industrial software testing,'' in \emph{Software
  Testing, Verification and Validation Workshops (ICSTW), 2014 IEEE Seventh
  International Conference on}.\hskip 1em plus 0.5em minus 0.4em\relax IEEE,
  2014, pp. 343--348.

\bibitem{Amalfitano2015}
D.~Amalfitano, A.~R. Fasolino, P.~Tramontana, B.~D. Ta, and A.~M. Memon,
  ``Mobiguitar: Automated model-based testing of mobile apps,'' \emph{IEEE
  Software}, vol.~32, no.~5, pp. 53--59, Sept 2015.

\bibitem{yang2013testing}
S.~Yang, D.~Yan, and A.~Rountev, ``Testing for poor responsiveness in android
  applications,'' in \emph{Engineering of Mobile-Enabled Systems (MOBS), 2013
  1st International Workshop on the}.\hskip 1em plus 0.5em minus 0.4em\relax
  IEEE, 2013, pp. 1--6.

\bibitem{fuchs2009scandroid}
A.~P. Fuchs, A.~Chaudhuri, and J.~S. Foster, ``Scandroid: Automated security
  certification of android,'' Tech. Rep., 2009.

\bibitem{feng2014apposcopy}
Y.~Feng, S.~Anand, I.~Dillig, and A.~Aiken, ``Apposcopy: Semantics-based
  detection of android malware through static analysis,'' in \emph{Proceedings
  of the 22nd ACM SIGSOFT International Symposium on Foundations of Software
  Engineering}.\hskip 1em plus 0.5em minus 0.4em\relax ACM, 2014, pp. 576--587.

\bibitem{huang2014asdroid}
J.~Huang, X.~Zhang, L.~Tan, P.~Wang, and B.~Liang, ``Asdroid: Detecting
  stealthy behaviors in android applications by user interface and program
  behavior contradiction,'' in \emph{Proceedings of the 36th International
  Conference on Software Engineering}.\hskip 1em plus 0.5em minus 0.4em\relax
  ACM, 2014, pp. 1036--1046.

\bibitem{octeau2013effective}
D.~Octeau, P.~McDaniel, S.~Jha, A.~Bartel, E.~Bodden, J.~Klein, and
  Y.~Le~Traon, ``Effective inter-component communication mapping in android
  with epicc: An essential step towards holistic security analysis,''
  \emph{USENIX Security}, 2013.

\bibitem{kleinberg2006algorithm}
J.~Kleinberg and E.~Tardos, \emph{Algorithm design}.\hskip 1em plus 0.5em minus
  0.4em\relax Pearson Education India, 2006.

\bibitem{jensen2013automated}
C.~S. Jensen, M.~R. Prasad, and A.~M{\o}ller, ``Automated testing with targeted
  event sequence generation,'' in \emph{Proceedings of the 2013 International
  Symposium on Software Testing and Analysis}.\hskip 1em plus 0.5em minus
  0.4em\relax ACM, 2013, pp. 67--77.

\bibitem{banerjee2014detecting}
A.~Banerjee, L.~K. Chong, S.~Chattopadhyay, and A.~Roychoudhury, ``Detecting
  energy bugs and hotspots in mobile apps,'' in \emph{Proceedings of the 22nd
  ACM SIGSOFT International Symposium on Foundations of Software
  Engineering}.\hskip 1em plus 0.5em minus 0.4em\relax ACM, 2014, pp. 588--598.

\bibitem{GATORWebSite}
\BIBentryALTinterwordspacing
Gator: Program analysis toolkit for android. [Online; accessed 5-October-2017].
  [Online]. Available:
  \url{http://web.cse.ohio-state.edu/presto/software/gator/}
\BIBentrySTDinterwordspacing

\bibitem{choi2013guided}
W.~Choi, G.~Necula, and K.~Sen, ``Guided gui testing of android apps with
  minimal restart and approximate learning,'' in \emph{Acm Sigplan Notices},
  vol.~48, no.~10.\hskip 1em plus 0.5em minus 0.4em\relax ACM, 2013, pp.
  623--640.

\bibitem{deng2017mutation}
L.~Deng, J.~Offutt, P.~Ammann, and N.~Mirzaei, ``Mutation operators for testing
  android apps,'' \emph{Information and Software Technology}, vol.~81, pp.
  154--168, 2017.

\bibitem{inozemtseva2014coverage}
L.~Inozemtseva and R.~Holmes, ``Coverage is not strongly correlated with test
  suite effectiveness,'' in \emph{Proceedings of the 36th International
  Conference on Software Engineering}.\hskip 1em plus 0.5em minus 0.4em\relax
  ACM, 2014, pp. 435--445.

\bibitem{ma2006mujava}
Y.-S. Ma, J.~Offutt, and Y.-R. Kwon, ``Mujava: a mutation system for java,'' in
  \emph{Proceedings of the 28th international conference on Software
  engineering}.\hskip 1em plus 0.5em minus 0.4em\relax ACM, 2006, pp. 827--830.

\bibitem{nilsson2007automated}
R.~Nilsson and J.~Offutt, ``Automated testing of timeliness: A case study,'' in
  \emph{Proceedings of the Second International Workshop on Automation of
  Software Test}.\hskip 1em plus 0.5em minus 0.4em\relax IEEE Computer Society,
  2007, p.~11.

\bibitem{weiMudroid}
Y.~Wei, ``Mudroid: Mutation testing for android apps,'' 2015.

\bibitem{zaeem2014automated}
R.~N. Zaeem, M.~R. Prasad, and S.~Khurshid, ``Automated generation of oracles
  for testing user-interaction features of mobile apps,'' in \emph{Software
  Testing, Verification and Validation (ICST), 2014 IEEE Seventh International
  Conference on}.\hskip 1em plus 0.5em minus 0.4em\relax IEEE, 2014, pp.
  183--192.

\bibitem{machiry2013dynodroid}
A.~Machiry, R.~Tahiliani, and M.~Naik, ``Dynodroid: An input generation system
  for android apps,'' in \emph{Proceedings of the 2013 9th Joint Meeting on
  Foundations of Software Engineering}.\hskip 1em plus 0.5em minus 0.4em\relax
  ACM, 2013, pp. 224--234.

\bibitem{hao2014puma}
S.~Hao, B.~Liu, S.~Nath, W.~G. Halfond, and R.~Govindan, ``Puma: programmable
  ui-automation for large-scale dynamic analysis of mobile apps,'' in
  \emph{Proceedings of the 12th annual international conference on Mobile
  systems, applications, and services}.\hskip 1em plus 0.5em minus 0.4em\relax
  ACM, 2014, pp. 204--217.

\bibitem{moran2018automated}
K.~Moran, B.~Li, C.~Bernal-C{\'a}rdenas, D.~Jelf, and D.~Poshyvanyk,
  ``Automated reporting of gui design violations for mobile apps,'' \emph{arXiv
  preprint arXiv:1802.04732}, 2018.

\bibitem{WALAframeworkWebSite}
\BIBentryALTinterwordspacing
IBM. Wala framework. [Online; accessed 12-April-2018]. [Online]. Available:
  \url{http://wala.sourceforge.net/wiki/index.php/Main_Page}
\BIBentrySTDinterwordspacing

\bibitem{ScanDroidWebSite}
\BIBentryALTinterwordspacing
Scandroid. [Online; accessed 15-April-2018]. [Online]. Available:
  \url{https://github.com/SCanDroid/SCanDroid}
\BIBentrySTDinterwordspacing

\bibitem{TroydWebSite}
\BIBentryALTinterwordspacing
Troyd. [Online; accessed 12-April-2018]. [Online]. Available:
  \url{https://github.com/plum-umd/troyd}
\BIBentrySTDinterwordspacing

\end{thebibliography}
\end{document}